\pdfoutput=1 % arXiv admin added this line

\documentclass[preprint,showpacs,preprintnumbers,amsmath,amssymb]{revtex4}

\usepackage{graphicx}% Include figure files
\usepackage{dcolumn}% Align table columns on decimal point
\usepackage{bm}% bold math
\usepackage{quantum}

\def\sigz{\op{\sigma_z}}
\def\sigm{\op{\sigma_-}}

\def\Dalpha[#1]{D_\alpha\left(#1\right)}

\def\Yb{$^{171}$Yb$^+$~}
\def\Dtwo{$^2$D$_{3/2}$~}
\def\Dthree{$^3$[3/2]$_{1/2}$~}

\def\bk{\hat{\textbf{b}}_\textbf{k}}
\def\bkdag{\hat{\textbf{b}}_\textbf{k}^\dagger}
\def\d{\textbf{d}}
\def\v{\hat{\textbf{v}}}
\def\vdag{\hat{\textbf{v}}^\dagger}

\def\k{\textbf{k}}

\def\r{\textbf{r}}

\def\nbar{\bar{n}}
\def\E{\overrightarrow{\textbf{E}}}

\def\Dx{\Delta\phi}
\def\Dy{\Delta\theta}
\def\sigm{\sigma_-}

\def\eo{\epsilon_0}
\def\uo{\mu_0}
\def\B{\textbf{B}}
\def\E{\textbf{E}}

\def\bw{\textbf{b}_\omega}
\def\avb{\langle \beta \rangle}
\def\avbmag{\langle |\beta|^2 \rangle}
\def\ld{\eta}
\def\wup{\omega_\uparrow}
\def\wdown{\omega_\downarrow}
\def\Lor[#1,#2]{\mathcal{L}( #1 , #2 )}
\def\xhat{\hat{\textbf{x}}}
\def\khat{\hat{\textbf{k}}}
\def\zhat{\hat{\textbf{z}}}

\begin{document}

\preprint{APS/123-QED}

\title{Protocol for Hybrid Entanglement Between a Trapped Atom and a Semiconductor Quantum Dot}

\author{Edo Waks}
% \homepage{http://www.Second.institution.edu/~Charlie.Author}
\affiliation{
University of Maryland\\
College Park, MD 20742
}%

\author{Christopher Monroe}
% \homepage{http://www.Second.institution.edu/~Charlie.Author}
\affiliation{
University of Maryland\\
College Park, MD 20742
}%

\date{\today}% It is always \today, today,
             %  but any date may be explicitly specified

\begin{abstract}
We propose a quantum optical interface between an atomic and solid state
system. We show that quantum states in a single trapped atom can be entangled
with the states of a semiconductor quantum dot through their common interaction
with a classical laser field. The interference and detection of the resulting
scattered photons can then herald the entanglement of the disparate atomic and
solid-state quantum bits. We develop a protocol that can succeed despite a
significant mismatch in the radiative characteristics of the two matter-based
qubits. We study in detail a particular case of this interface applied to a
single trapped \Yb ion and a cavity-coupled InGaAs semiconductor quantum dot.
Entanglement fidelity and success rates are found to be robust to a broad range
of experimental nonideal effects such as dispersion mismatch, atom recoil, and
multi-photon scattering.  We conclude that it should be possible to produce
highly entangled states of these complementary qubit systems under realistic
experimental conditions.
\end{abstract}

\pacs{Valid PACS appear here}% PACS, the Physics and Astronomy
                             % Classification Scheme.
%\keywords{Suggested keywords}%Use showkeys class option if keyword
                              %display desired
\maketitle

%\section{\label{sec:Introduction} Introduction}

%\input{Introduction}

\section{Introduction}

Quantum entanglement, long considered to be the most puzzling aspect of quantum
mechanics \cite{EPR, Schrodinger}, is now realized to be a potential resource
for enhanced processing and communication of information.  The field of quantum
information science exploits quantum entanglement for tasks that are otherwise
impossible or inefficient using conventional information processing approaches
\cite{nielsen}. Recent advances in the control of physical systems, ranging
from isolated atoms and photons to individual degrees of freedom in condensed
matter, have shown great promise in the development of quantum information
hardware \cite{PW08}.

The majority of work to date has been centered around entanglement of identical
quantum systems such as identical atoms and photons. There is great interest in
extending entanglement over disparate quantum systems. Such hybrid entanglement
can exploit advantages of each individual system to enhance capabilities of
quantum technology.  An important example is the hybrid entanglement between
matter and photonic quantum systems.  This type of entanglement, which has been
demonstrated using both trapped ions~\cite{Moehring2004} and atomic
ensembles~\cite{Sherson06}, enables one to combine the advantages of the long
coherence times of atomic systems with the ability of photonic systems to
transport quantum information.  Another important example is the recent
entanglement of two different species of trapped atomic ions, where one species
(Al$^+$) has excellent coherence properties and the other (Be$^+$) allows
efficient qubit measurement \cite{Schmidt05}.

Here, we theoretically investigate the possibility of creating hybrid
entanglement between semiconductor and atomic quantum systems. Specifically, we
propose a protocol for entangling a quantum dot (QD) in a microcavity with a
trapped atom through a common photonic interaction. Such hybrid entanglement is
expected to stimulate new concepts for distributed quantum computation that
exploit the long coherence times of trapped ions with the fast dynamics and
strong atom-photon interactions of a QD. We show that a common photonic channel
can link these disparate systems to achieve hybrid-matter quantum entanglement
despite significant mismatches between atomic and semiconductor qubits.

Photon mediated entanglement has been proposed for entangling like systems such
as atoms~\cite{DLCZ01, Sherson06}, NV centers in diamond \cite{Childress06},
and cavity-coupled QDs~\cite{WaksVuckovicPRL2006,SridharanWaks2008}. It has
been experimentally demonstrated between electrostatically trapped
ions~\cite{Moehring07} and atomic
ensembles~\cite{ChouRiedmatten2005,ChouJulien2007,ChoiDeng2008}.  To extend
such ideas to entangle different matter qubits requires a protocol that is
robust to significant mismatch in spectral and temporal properties.  In this
paper we propose such a protocol in which the atomic system is coupled to a
free-space field by elastic scattering, while the QD is coupled via cavity
interaction. We extensively investigate a particularly promising implementation
of this protocol in which an electrostatically trapped \Yb~ion is entangled
with a cavity coupled indium arsenide (InAs) QD.  We address the central
challenges in realizing this hybrid quantum link, including the mismatch in the
optical spectra of these systems and decoherence processes that are particular
to each node of the hybrid circuit. In particular, we analyze in detail
``which-path" decoherence due to atom recoil and multi-photon scattering and
derive an analytical expression for entanglement fidelity and entanglement
rate.

In section~\ref{sec:Protocol}, we give a low-level description of the proposed
entanglement protocol.  Section~\ref{sec:ScatterCalc} provides detailed
calculations of the atom scattering amplitudes, and analyzes the effect of
dispersion. Section~\ref{sec:Recoil} considers atom motion and recoil, and
calculates the fidelity under realistic experimental conditions. In
section~\ref{sec:WhichPath} we consider the effects of leakage of which-path
information due to multi-photon scattering.  The final section investigates the
validity of the weak excitation approximation which is extensively used to
calculate both atomic and scattering and reflection from the cavity-QD system.

\section{Basic Protocol} \label{sec:Protocol}

\begin{figure}
\centering
\includegraphics[width=6 in]{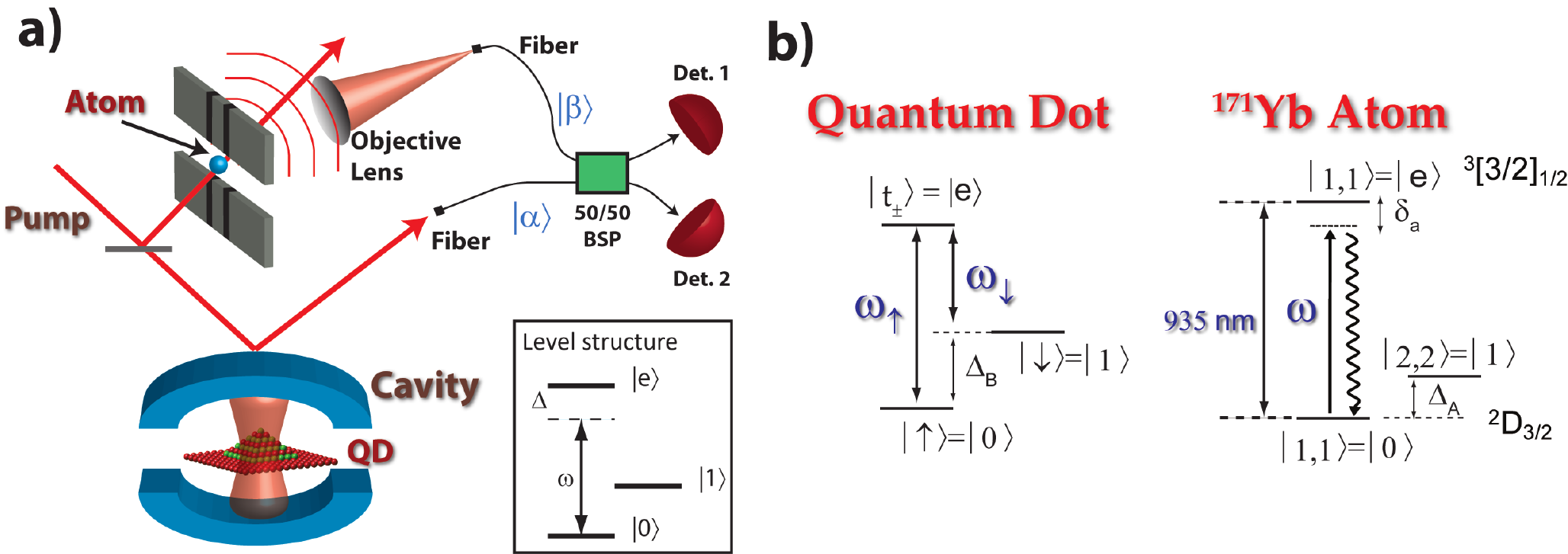}
\caption{ \textbf{(a)} Proposed setup for entangling an atom with a QD. A
trapped atom and a cavity-coupled QD are held at different
spatial locations. Both QD and atom have a level structure shown in the
inset, and a common pump beam excites both systems. The reflected light from the
cavity is mixed with the scattered light from the atom on a $50/50$ beamsplitter.
Path lengths are set for constructive interference at detector 1.  A detection event
at detector 2 places the two systems in an entangled state.  \textbf{(b)} Specific
level structure for the QD and atom.  The spin states of a singly charged QD
can serve to create the desired qubit states, while the positive or negative
trion state ($\|t_\pm>$)serves to couple the qubit to the cavity. The two qubit states
are split in energy by $\Delta_B$ due to an applied magnetic field. For the atom we consider
the specific example of a \Yb ion and use
the hyperfine $\|1,1>$ and $\|2,2>$ states (with hyperfine splitting $\Delta_A=0.86$ GHz)
in the \Dtwo manifold to store
quantum information, while optical transition to the $\|1,1>$ state in the
\Dthree manifold results in elastic scattering of the pump field.  Selection
rules ensure that scattering will occur only when the atom is in state
$\|2,1>$.}\label{fig:setup}
\end{figure}

Fig.~\ref{fig:setup}a illustrates the proposed method for entangling a QD with
an atom. The QD is coupled to a resonant microcavity while the atom is
electromagnetically confined at a remote location. Each system consists of a
qubit represented by the two states, $\|0>$ and $\|1>$, where state $\|0>$ is
coupled to a third excited state $\|e>$ via an optical transition. We consider
the case where this transition in the QD and atom occur at similar (not
necessarily identical) optical wavelengths. The entanglement method we consider
is general, and can apply to a broad range of atomic and semiconductor systems.
However, to perform calculations we specialize to the case of an InGaAs QD that
is entangled with an electrostatically trapped \Yb ion.  The assumed level
structure for these two qubit systems is shown in Fig.~\ref{fig:setup}b.  We
focus on \Yb because of its optical \Dtwo to \Dthree transition, which occurs
at 935 nm and is therefore compatible with the near infrared transition
wavelength of InGaAs QDs.  For the QD, the qubit states can reside in the
Zeeman split spin states of a single charge carrier, with an optical coupling
to an excited trion state, as recently
demonstrated~\cite{XuSun2008,PressLadd2008}. The loading of a single spin into
a microcavity coupled QD has also been recently achieved~\cite{rakher:097403}.
For the atom, the qubit states are represented by long-lived electronic/nuclear
hyperfine states, with an optical coupling to a dipole-allowed excited
electronic state~\cite{WinelandBlatt08}. We consider the hyperfine qubit states
residing in the metastable \Dtwo level, with a coupling to the excited \Dthree
level at 935 nm \cite{Yb-qubit}.

Entanglement is established by first initializing the state of each of the two
qubits to superposition state $\|\psi_i> = (\|0>+\|1>)/\sqrt{2}$.  The QD
system can be initialized using experimentally demonstrated
single~\cite{FuClark08} and two-laser coherent control
techniques~\cite{PressLadd08,XuSun08}, while the atom can be controlled using
optical pumping and microwave or stimulated Raman transitions
\cite{WinelandBlatt08,Yb-qubit}. Following initialization, a laser pulse is
coherently split into two components directed to the two quantum systems. One
of the components is reflected off of the cavity containing a QD while the
second component drives the atom off resonance, resulting in an elastically
scattered field that is phase coherent with the input field. Phase coherence of
the elastically scattered component has been experimentally verified through
Young's interference~\cite{itano:1998}, and has also been theoretically
investigated~\cite{eichmann:1993}.  We define $\|\alpha>$ as the coherent field
reflected from the cavity, and $\|\beta>$ as the field elastically scattered
from the atom. The two fields are combined on a beamsplitter whose path lengths
are set for constructive interference at detector 1 (shown in
Fig.~\ref{fig:setup}a), requiring optical interferometric stability over the
system.  A detection event at detector 2 will collapse the state of the atom
and QD into an entangled state.

To understand the entanglement formation process, we first assume that both
fields are quasi monochromatic, phase coherent, and sufficiently weak that they
can be expanded to first order in photon number. In later sections we will
derive the entanglement fidelity under more realistic experimental conditions.
We define $\wup$ and $\wdown$ as the trion resonant frequencies for the two
different spin states of the charged QD, which are detuned by $\Delta_B$ due to
an applied external magnetic field. The cavity resonant frequency is defined as
$\omega_c$, and input field frequency as $\omega$. The reflection coefficient
$r$ and transmission coefficient $t$ of the cavity
are~\cite{WaksVuckovicPRL2006,ZhouXiang2005,HughesKamada2004}
\begin{eqnarray}
r(\omega) &=& \frac{-i\Delta + \mathcal{C}\mathcal{L}(\delta_{qd},\gamma_{qd})}
                   {1-i\Delta+\mathcal{C}\mathcal{L}(\delta_{qd},\gamma_{qd})} \\ \label{eq:r}
t(\omega) &=& \frac{1}
                   {1-i\Delta+\mathcal{C}\mathcal{L}(\delta_{qd},\gamma_{qd})} \label{eq:t}
\end{eqnarray}
where $\delta_{qd} = \omega - \omega_{\uparrow,\downarrow}$ is the detuning
between the cavity and the QD (which depends on the spin state of the QD), and
$\Delta = (\omega-\omega_c)/\kappa$ is the laser-cavity detuning scaled to the
cavity linewidth $\kappa$.   The QD exciton decay rate is given by
$\gamma_{qd}$, $\mathcal{C}= 4g^2/(\gamma_{qd}\kappa)$ is the QD-cavity
cooperativity, $g$ is that QD-cavity coupling strength, and
$\mathcal{L}(\delta,\gamma)=\gamma/(\gamma-i\delta)$ is a Lorentzian
profile~\cite{WaksVuckovicPRL2006}.  The amplitude of the reflected field is
given by $\alpha=\alpha_{in} r(\omega)$, where $\alpha_{in}$ is the incident
field amplitude.

The incident field is set to be resonant with the cavity ($\omega=\omega_c$) so
that $\Delta=0$. We also set $\wup=\omega_c$ so that the QD is resonantly
coupled to the cavity mode when it is in state $\|0>=\|\uparrow>$.  In this
case we have $\delta_{qd}=0$, and in the limit of large atomic cooperativity
$|\mathcal{C}|\gg 1$, the cavity reflectivity approaches $r(\omega_c)=1$.  If
the QD is instead in state $\|1>=\|\downarrow>$ there is little coupling
between the QD and cavity either due to selection rules ($g=0$) or large
detuning ($\delta_{qd} \gg g^2/\kappa$). Thus,
$\mathcal{C}\Lor[\delta_{qd},\gamma_{qd}]\to 0$ and $t(\omega_c) \to 1$, so all
of the light is transmitted through the cavity. Therefore, the qubit state of
the QD will switch the cavity from being highly reflecting to highly
transmitting. This operation can be viewed as a controlled-NOT gate between the
state of the QD and the propagation direction of the scattered light.
Controlled reflectivity of a cavity via a single QD has been reported in
several works~\cite{EnglundFaraon07,SrinivasanPainter2007,rakher:097403}.

For the atom, we assume that the driving field is detuned from the resonant
transition frequency $\omega_a$ by $\delta_a =\omega-\omega_a$. If the atom is
in state $\|0>$, it will induce off-axis elastic scattering via the near
resonant optical transition to state $\|e>$.  The scattered field is collected
by an objective lens and coupled into a single mode fiber. We define $\beta$ as
the coherent state amplitude of the field scattered into the fiber. In
contrast, when the atom is in state $\|1>$ it will not scatter any light since
it cannot make an optical transition due to selection rules. For the case of
\Yb, we identify $\|0>$ and $\|1>$ states as the $F=1, m_F=1$ and $F=2, m_F=2$
hyperfine levels of the metastable ($53$ ms lifetime) electronic \Dtwo state,
with a frequency splitting of $0.86$ GHz. With the $935$ nm input laser field
linearly polarized parallel to a weak magnetic field, the $\|0>$ state couples
to the \Dthree $F=1, m_F=1$ hyperfine level we identify as state $\|e>$, while
the $\|1>$ state remains dark, as indicated in Fig.~\ref{fig:setup}b.

Under the assumption that the reflected field from the cavity and scattered
field from the atom are sufficiently small so that they can be expanded to
first order in photon number, a detection event at detector $2$ will collapses
the state of the QD and atom onto
\begin{equation}
\|\psi_f> = D^{-1} [(\alpha -\beta)\|0>_{qd}\|0>_a + \alpha \|1>_{qd}\|0>_a -
\beta\|0>_{qd}\|1>_a]
\end{equation}
where $D^2= (|\alpha - \beta|^2+|\alpha|^2 +|\beta|^2)$. By tuning the phase
and amplitude of the input fields so that $\alpha = \beta$, the above state
becomes a maximally entangled Bell state $\|\psi_f>=(\|1>_{qd}\|0>_a -
\|0>_{qd}\|1>_a)/\sqrt{2}$. It should be noted that, so long as the
quasi-monochromatic limit is valid, a perfect entangled state can be generated
even when the QD and atom have different resonant frequencies and decay rates.

\section{Fidelity under pulsed excitation} \label{sec:ScatterCalc}

The ideal protocol described in section~\ref{sec:Protocol} considers the
quasi-monochromatic limit, where the input pulses are sufficiently long in time
to be considered as single frequency.  In a real experiment, the entanglement
operation must be completed before the QD and atom have had time to decohere,
which requires excitation with short optical pulses. The coherence time of the
atomic hyperfine states is long (53 ms), but the QD spin coherence time is much
shorter (10 ns) due to hyperfine interactions with nuclear
spin~\cite{DuttCheng2005}. Recent progress in optical locking of nuclear spin
polarization suggests that much longer lifetimes in the microsecond regime may
be possible~\cite{XuYao2009}. Nevertheless, it is important to consider the
effect of short pulses ($0.1-10$ ns) on the fidelity of the generated entangled
state.

Because we consider the weak excitation regime, the cavity-QD system and
trapped atom are linear scatterers. Therefore, we may analyze the effect of
time varying excitation by fourier decomposing the time dependent input fields
and looking at the scattering behavior of each fourier component independently.
Such approach would not be valid in the strong excitation limit where nonlinear
scattering occurs due to absorption saturation.  We defer that discussion to
section~\ref{sec:nonlinear}.

%The scattering amplitude for the cavity-QD system is fully characterized by
%Eq.~\ref{eq:r} and Eq.~\ref{eq:t}.

In order to analyze the spectral properties of the field scattered by the atom,
we need a more precise expression for its scattering amplitude. To derive this
amplitude, we begin with the standard system Hamiltonian for the atom given by
 \begin{eqnarray*}\label{eq:H}
    \textbf{H} & = & \sum_\k \hbar \bkdag\bk + \frac{\sigma_z}{2}\omega_a +
    \sum_\k \hbar g_\k \left(\sigma_+\bk e^{i\k\cdot\r} +
    \sigma_-\bkdag e^{-i\k\cdot\r}\right)  + \\
    && \hbar \int d\omega \left[\Omega(\omega)\sigma_+ e^{i(\k_\omega\cdot\r-\omega t)} +
    \Omega^*(\omega)\sigma_- e^{-i(\k_\omega\cdot\r+\omega t)}\right] + V_{trap}
  \end{eqnarray*}
In the above equation $\bk$ is a bosonic operator for a free space mode
obtained using box boundary conditions, $\omega_a$ is the resonant frequency of
the atom, $\sigma_z$ is the population difference operator for atom scattering
transition, and $\sigma_-$ is the transition lowering operator. The scattered
modes are coupled to the atomic dipole by the coupling strength
$g_\k=-\d\cdot\epsilon_\k\sqrt{\omega_\k/\hbar\epsilon_0 V}$ where $V$ is the
quantization volume and $\d = d \zhat$ is the atomic dipole moment. We assume
that the input field is sufficiently bright so that it may be treated as a set
of classical Rabi frequencies $\Omega(\omega)$. The input field is assumed to
be largely collimated and propagating along the x-axis so $\k_\omega = \xhat
\omega/c$. The potential $V_{trap}$ is the harmonic trapping potential of the
atom, while $\r$ is the location of the atom in the trap.

We first consider the stationary atom limit, where $\r$ is a real vector.
Atomic recoil, which requires us to account for the quantized operator nature
of $\r$, will be addressed in the next section. When the atom is in qubit state
$\|0>$ the Heisenberg equations of motion for the atomic and cavity field
operators are given by
  \begin{eqnarray}
    \frac{d\bk}{dt} & = & -i\omega_\k \bk - i g_\k \sigm e^{i\k\cdot\r} \\ \label{eq:bk}
    \frac{d\sigm}{dt} & = & -(i\omega_a + \gamma_a) - i \int d\omega \Omega(\omega) e^{i(\k\cdot\r - \omega t)} \label{eq:sigm}
  \end{eqnarray}
where we have added the decay term $\gamma_a$ which accounts for the total
dipole decay, including dephasing.  In Eq.~\ref{eq:sigm} we assume that the
scattered field is very weak as compared to the strong classical field
$\Omega$, so its driving term is ignored in the equation.  We also make the
weak excitation approximation that the atom remains mostly in the ground state,
so $\sigz\to -1$~\cite{WaksVuckovicPRL2006}.

Eq.~\ref{eq:sigm} can be solved by direct integration to give
  \begin{equation} \label{eq:sigmamp}
    \sigm = -i\frac{\Lor[\delta_a,\gamma_a]}{\gamma_a}\int_{-\infty}^{\infty} d\omega \Omega(\omega) e^{i(\k_\omega\cdot\r - \omega t)}
  \end{equation}
where $\delta_a = \omega - \omega_a$ and, as before, $\Lor[\omega,\gamma] =
\gamma/(i\omega +\gamma)$. In deriving the above equation we consider only the
forced response of the atom driven by the classical Rabi frequency, and ignore
the transient response due to turn-on of the input field.  Plugging into
Eq.~\ref{eq:bk} and integrating, we subsequently obtain
  \begin{equation} \label{eq:bksolved}
    \bk = -\pi e^{-i\omega_\k t} \frac{g_\k \Lor[\delta_a,\gamma_a]}{\gamma_a}\Omega{\omega_\k} e^{i(\k_\omega - \k)\cdot\r}
  \end{equation}
The classical pump field that drives the atom is derived from a laser source
and is therefore a coherent state.  The atom is a linear scatterer so the
scattered field must also be a coherent state whose quantum state is fully
characterized by the scattered amplitude
  \begin{equation}\label{eq:ScatAmp}
    \beta_\k= \langle \bk \rangle = -\pi \frac{\Lor[\omega_a,\gamma_a]}{\gamma_a} \Omega(\omega_k) e^{i(\k_\omega - \k)\cdot\r}
  \end{equation}
In Appendix~\ref{app:NumScat} we show that the total number of scattered
photons $N_s=\sum_\k |\beta_k|^2$ is given by
  \begin{equation} \label{eq:NumScat}
    N_s = \left|\Lor[\omega_a,\gamma_a]\right|^2 \left(\frac{\gamma_r}{\gamma_a}\right)^2\sigma_0 n_i
  \end{equation}
where $\omega_0$ is the center frequency of the incident field, $n_i$ is the
input photon flux density given by $n_i = I/\hbar\omega_0$ where $I$ is the
input field intensity, $\gamma_r$ is the radiative decay rate of the atom to
state $\|0>$, and $\sigma_0=3\lambda^2/2\pi$ is the atomic cross section.

  \begin{figure}
  \centering
  \includegraphics[width=5in]{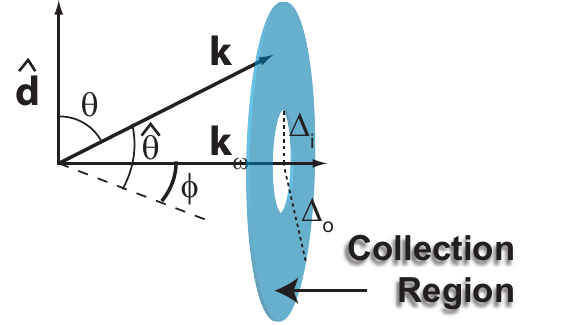}
  \caption{Model for collection optics.  All k-vectors between the angles $\Delta\theta_i$
  and $\Delta\theta_o$ from the input propagation direction (shaded region) are collected by the fiber.  All other k-vectors are
  not collected}
  \label{fig:collection}
  \end{figure}

The important quantity for our analysis is $\beta (\omega)$, the amplitude of
the coherent field coupled to the fiber at frequency $\omega$. Since the
collection optics are comprised of only linear optical elements, the fiber
amplitude $\beta(\omega)$ is related to the free space scattered amplitudes
$\beta_\k$ by the linear transformation
  \begin{equation}
    \beta(\omega) = \sum_{|\k|=\omega/c} s_\k \beta_\k
  \end{equation}
where $s_\k$ are a set of complex scattering coefficients representing a
unitary transformation.  To proceed, we need a model for the collection optics.
Here we consider a simple model, illustrated in Fig.~\ref{fig:collection},
where all k-vectors between a solid angle $\Delta_i$ and $\Delta_o$ (taken with
respect to the propagation direction of the pump) are collected by the fiber.
The remaining k-vectors do not couple to the fiber. Thus,  $s_\k = s$ if $\k$
is within the collection window, and $s_\k = 0$ if $\k$ is outside the
collected solid angle. The omission of solid angles between 0 and $\Delta_i$ is
included in order to reject the input beam. For a well collimate input, we can
take the limit that $\Delta_i\to 0$. The simple model we consider captures all
of the relevant physics, and provides good numerical accuracy for our
calculations.  A more accurate model would treat $s_\k$ as a gaussian
transverse distribution, matching the gaussian profile of the single mode
fiber.  This model would significantly complicate the calculations but would
ultimately yield similar results.

Our calculations focus on the \textit{paraxial limit}, where the collected
solid angle $\Delta_o$ is small, so that the collected light propagates nearly
parallel to the input beam.  We focus on this range of collection angles
because it is known to minimize the effect of atomic
recoil~\cite{itano:1998,eichmann:1993}, as we will analyze in more detail in
the next section. Using the simple model for the collection optics, we shown in
Appendix~\ref{ap:FiberAmp} that in the paraxial limit
  \begin{equation}\label{eq:beta}
    \beta(\omega) = \beta_0 \Lor[\delta_a,\gamma_a] \Omega(\omega)
  \end{equation}
with
  \begin{equation}\label{eq:beta0}
    \beta_0 = \frac{d \omega_0^{3/2} }{2c\gamma_a \sqrt{2\eo\hbar L_x A}} \int_A d\theta d\phi e^{i k_0(\xhat - \khat)\cdot\r}
  \end{equation}
In the above equation the integral is taken over the collection area, while
$A=2\pi (\cos\Delta_i - \cos\Delta_o) \approx \pi (\Delta_o^2 - \Delta_i^2)$ is
the area that the collection region occupies on the unit sphere and $L_x$ is
the length of the quantization volume in the $x$ direction. For simplicity we
assume that $\omega_k$ can be replaced by its average value $\omega_0$ and
$k_0=\omega_0/c$.  We do not make this substitution in the Lorentzian function,
however, because near resonance this function will vary rapidly even over a
narrow bandwidth of interest.

In the monochromatic limit, we could always achieve the matching condition
$\alpha=\beta$ by changing the amplitude and phase of the incident fields.  But
under pulsed excitation, the reflection coefficient of the QD-cavity system has
a spectral profile described by Eq.~\ref{eq:r}, while the atom scattering
amplitude follows primarily a Lorentzian profile. Thus, each spectral component
will require a different matching condition.  Since we cannot satisfy the
matching condition perfectly for each frequency we expect that the generated
state will no longer be perfectly entangled. It has previously been shown that
the spectral width over which high reflection is achieved when a QD is coupled
to a cavity is given by the inverse modified spontaneous emission lifetime of
the cavity enhanced QD transition~\cite{WaksVuckovicPRL2006}, typically 10-50
GHz. This bandwidth represents a response time which is much faster than any
decay rates of the atom.  Thus, we expect the fidelity of the generated
entangled state to be dominated by the dispersive properties of the atom, which
are much narrower in frequency the those of the QD.

To calculate the fidelity of the generated entangled state under pulsed
excitation, we begin with the state of the system after the field from the
QD-cavity system is mixed with the scattered field from the atom on the
beamsplitter, as shown in Fig.~\ref{fig:setup}.  We once again assume that the
scattered fields are sufficiently weak so that they may be expanded to first
order in photon number.  After the beamsplitter the state of the fields, atom
and QD can be written as
  \begin{equation}
    \|\psi_f> = \int d\omega \frac{1}{2\sqrt{2}} \left[\int \left(\alpha(\omega) -\beta(\omega)\right)\|00> + \alpha(\omega)\|01> -\beta(\omega)\|10> \right]\vdag_\omega\|0> + \|f>
  \end{equation}
In the above state the operator $\v_\omega$ is the bosonic operator for a
photon in the mode detected by detector 2, and state $\|f>$ is an un-normalized
state representing the remaining components of the wavefunction that do not
contain any photons in the detection mode. From the expressions for the
reflection coefficients we have
  \begin{eqnarray}
    \alpha(\omega) & = & \alpha_0 r(\omega) \Omega(\omega) \\
    \beta(\omega) & = & \beta_0 \Lor[\delta_a,\gamma_a] \Omega(\omega)
  \end{eqnarray}
where $\alpha_0$ and $\beta_0$ are complex amplitudes that can be adjusted
 by selecting the amplitude and phase of the input pulse. In the
ideal case we would have $\alpha(\omega)=\beta(\omega)$ for all frequencies,
which would reproduce the ideal fidelity of the monochromatic limit.
Unfortunately, dispersion prohibits us from achieving this for all values of
$\omega$.  The best we can do is adjust the amplitudes so that
$\alpha(\omega_0)=\beta(\omega_0)$, where $\omega_0$ is the center frequency of
the pulse.

The fidelity of the entangled state can be determined by first calculating the
reduced density matrix of the QD-atom system conditioned on a detection event
at detector 2.  This reduced density matrix is given by
  \begin{equation} \label{eq:FidSpectrum}
    \rho = \frac{ \textbf{Tr}_{fields} \{ \textbf{P}\|\psi_f>\<\psi_f|\}}{\textbf{Tr}\{\textbf{P} \|\psi_f>\<\psi_f|\}}
  \end{equation}
where $\textbf{P}$ is a projector onto the subspace containing at least one
photon in the detection mode for detector 2.  The fidelity can be calculated
using $F=\<\psi_-|\rho\|\psi_->$, where $\|\psi_->$ is the ideal spin singlet
entangled state.  The expression for fidelity is given by
 \begin{equation}\label{eq:Fspectral}
    F = \frac{1}{4} \frac{\int d\omega \left| \alpha(\omega) +\beta(\omega) \right|^2}{\int d\omega \left| \alpha(\omega)\right|^2 +\left|\beta(\omega) \right|^2-\textbf{Re}\{\alpha(\omega)\beta(\omega)\}}
  \end{equation}

  \begin{figure}
  \centering
  \includegraphics[width=5in]{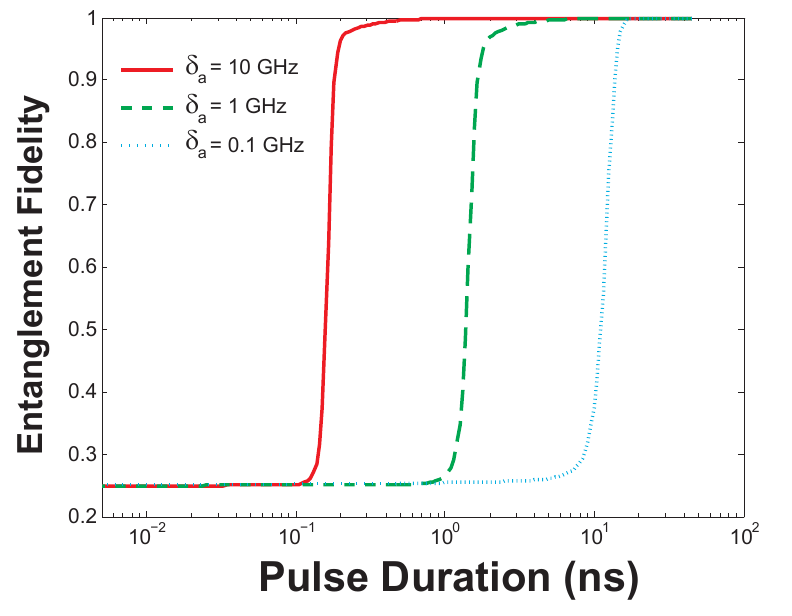}
  \caption{Entanglement fidelity as a function of pulse duration for several different
values of atomic detuning $\delta_a = \omega-\omega_a$. For each detuning, there is a
certain value of the pulse duration where the fidelity quickly drops from 1
(ideal entangled state) to 0.25 (no coherence between $\alpha$ and $\beta$).
Larger $\delta_a$ enables shorter pulses before fidelity drops.}
  \label{fig:F_pulse}
  \end{figure}

Fig.~\ref{fig:F_pulse} plots the fidelity as a function of the pulse duration
of the external field for several values of the detuning from atomic resonance.
The curves are obtained by numerical integration of Eq.~\ref{eq:Fspectral}. We
assume the input pulse is gaussian with a spectrum given by
$\Omega(\omega)=\Omega_0e^{-\tau^2(\omega-\omega_0)^2/4}e^{i(\k\cdot\r-\omega
t)}$, where $\tau$ is the pulse duration.  To calculate the reflection
coefficient for the cavity containing the QD we use parameters that are
appropriate for InAs QDs coupled to photonic crystal defect cavities. Using
experimental values from recent work~\cite{EnglundFaraon07}, we set $g/2\pi =
16$ GHz, $\delta_{qd}=0$, $\kappa/2\pi =25$ GHz, and $\gamma_{qd}/2\pi = 1$
GHz. For the \Yb atomic ion we use $\gamma_{a}/2\pi = 4.2$ MHz, the linewidth
of the \Dthree state.

In the long pulse limit, the fidelity approaches its ideal value all atomic
detunings. However, for each value of atomic detuning the fidelity makes a
rapid transition from $F=1$ to $F=0.25$ at some critical pulse duration. Thus,
the range of pulses for which the monochromatic approximation is valid will
depend on $\delta_a$. At detuning of $\delta_a=0.1$ GHz, the transition occurs
roughly at 10 ns pulse duration, which is on the order of the coherence time of
the QD. By increasing the detuning to 1 GHz, the fidelity transitions at 1 ns
pulse duration, well below the QD decoherence time.  By further increasing the
detuning to 10 GHz it is possible to use 100 ps pulses. Thus, by increasing the
atomic detuning we can use shorter pulses to achieve high fidelity. This
tradeoff occurs because the dispersion, dominated by a Lorentzian function, is
maximum near resonance and tails off with larger detuning.

 \begin{figure}
  \centering
  \includegraphics[width=5in]{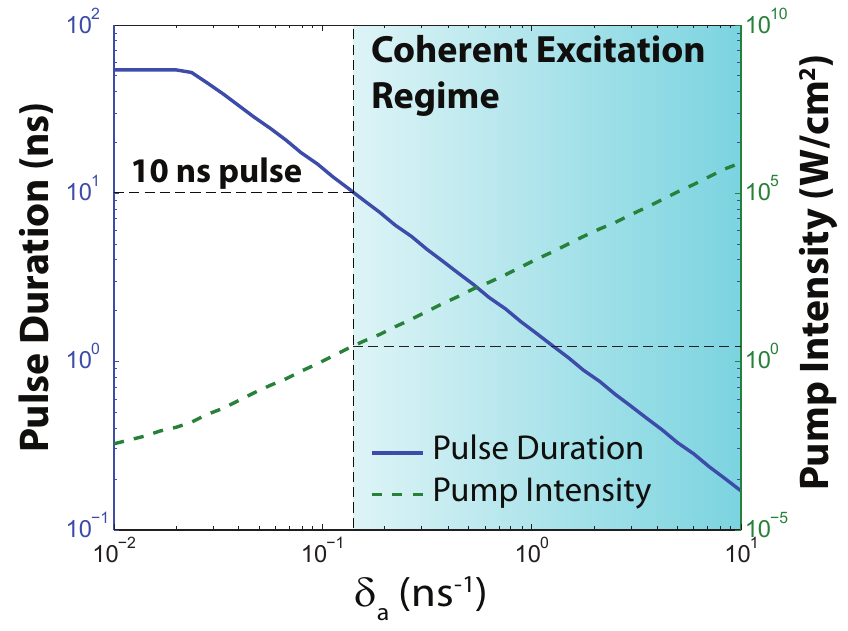}
  \caption{Solid line plots pulse duration required to achieve fidelity of 0.9
as a function of $\delta_a$. Dashed line plots pump intensity required to
achieve 0.1 scattered photons in the pulse. As $\Delta\omega_a$ increases, we
can use shorter pulses but require higher excitation energies to achieve the
same scattering rate. If the coherence time of the QD is 10 ns, the shaded
region represents the parameter regime where entanglement can be achieved.
}
  \label{fig:pulse_delta}
  \end{figure}

The disadvantage of going to larger detuning is that we need to use more pump
power in order to achieve the same scattering rate. The number of scattered
photons is given by Eq.~\ref{eq:NumScat}.   From this equation one can see that
as $\delta_a$ increases the Lorentzian decays in amplitude, forcing us to
increase the incident photon density $n_i$ to attain the same number of
scattered photons. The tradeoff between pulse duration and pump intensity is
shown in Fig.~\ref{fig:pulse_delta}. The panel plots both the pulse duration
required to achieve a fidelity of 0.9 (solid blue curve) and the pump intensity
required to scatter $N_s=0.1$ photons (dashed green curve) as a function of
$\delta_a$. The choice of $N_s=0.1$ ensures that the collected field is
sufficiently weak to be expanded to first order in photon number. The shaded
area, labeled the \textit{coherent excitation regime}, represents the operating
region where the pulse duration is less than 10 ns, the typical dephasing time
of the InAs QD electron spin \cite{PhysRevLett.94.047402}. The coherence time
of the atom is longer ($53$ ms), so the QD limits the coherence time of the
overall entangled state.  One can see from the figure that pump intensities as
low as $2$ W/cm$^2$ would be sufficient to enable the use of 10 ns pulses with
the specified scattering rate.

\section{Decoherence due to atom recoil} \label{sec:Recoil}

Entanglement between the atom and QD relies on the assumption one cannot
distinguish whether a photon was reflected from the cavity or scattered from
the atom, even in principle.  Atomic recoil can serve to betray this
''which-path'' information.  When the atom scatters a photon there is a
probability that it will recoil, leaving residual kinetic energy in the
motional degrees of freedom of the center-of-mass wavefunction.  To achieve
high fidelity we require that this recoil probability is small.

Atomic recoil is already present in the expression for the scattered field
amplitude given in Eq.~\ref{eq:beta} and Eq.~\ref{eq:beta0}.  Previously we
assumed that the position $\r$ of the atom was fixed, and therefore the
integral term in the expression was simply a complex constant.  To include the
effect of recoil, we must keep $\r$ as the position operator and trace it out
over the motional degrees of freedom of the atom. We define the initial state
of the atom-QD system by the density matrix
  \begin{equation}
    \rho_i = \sum_n p(n) \|\psi_{nf}>\<\psi_{nf}|
  \end{equation}
where
  \begin{equation} \label{eq:psin}
    \|\psi_{ni}> = \frac{1}{2}\left(\|00> + \|11> + \|10> + \|01>\right)\|n>\|vac>
  \end{equation}
In the above equations, $\|n>$ is the harmonic oscillator eigenstate for the
center of mass motion of the atom, $\|vac>$ is the vacuum field for all optical
modes, and $p(n)$ is assumed to be a thermal distribution.  After interaction
with the two input fields, the state in Eq.~\ref{eq:psin} is transformed into
  \begin{equation}
    \|\psi_{nf}> = \|\psi_d>\|n>\|v> + \|f_n>
  \end{equation}
with
  \begin{equation}
    \|\psi_d> = \frac{1}{2} \left[ \left(\frac{\alpha - \beta}{\sqrt{2}}\right)\|00> + \frac{\alpha}{\sqrt{2}}\|01> - \frac{\beta}{\sqrt{2}}\|10> \right]
  \end{equation}
The state $\|v>$ represents a single photon in the detection mode that is
monitored by detector 2, while $\|f_n>$ is once again a wavefunction orthogonal
to $\|v>$ representing the state of the QD, atom, and all fields when there are
no photons in mode $\v$. Conditioned on a detection event at detector 2, the
final density matrix of the system is given by
  \begin{equation}\label{eq:rhof}
    \rho_f = \frac{\textbf{Tr}_n\{\sum_n p(n) \|\psi_{nf}>\<\psi_{nf}|\}}{\textbf{Tr}\{\sum_n p(n) \|\psi_{nf}>\<\psi_{nf}|\}}
  \end{equation}
where $\textbf{Tr}_n$ represents a trace over all degrees of freedom except for
the qubit states of the atom and QD.

The fidelity $F$ of the final state can be defined as the overlap between the
actual state of the system and the desired state $\|\psi_-> = (\|01> -
\|10>)/\sqrt{2}$. Thus,
  \begin{equation}
    F = \<\psi_-|\rho_f\|\psi_->
  \end{equation}
Using the definition of $\rho_f$ in Eq.~\ref{eq:rhof} we attain the following
expression for the fidelity:
  \begin{equation}
    F = \frac{1}{4} \frac{|\alpha|^2 + \avbmag + 2\alpha\textbf{Re}\{\avb \}}{|\alpha|^2 + \avbmag - \alpha\textbf{Re}\{\avb \}}
  \end{equation}
where
  \begin{eqnarray}
    \avb & = & \sum_n p(n) \<n|\beta\|n> \\ \label{eq:avb}
    \avbmag & = & \sum_n p(n) \<n| |\beta|^2 \|n> \\ \label{eq:avbmag}
  \end{eqnarray}
The expressions in Eq.~\ref{eq:avb} and Eq.~\ref{eq:avbmag} can be evaluated
under the assumption that the ion occupies a thermal distribution.  In the
case, we can then write~\cite{itano:1998,Mermin1966,Bateman1992}
  \begin{equation} \label{eq:ThermAvg}
    \langle e^{i(\k-\k')\cdot\r} \rangle = e^{-\ld^2(1-\cos\hat{\theta}\cos\phi)(\nbar + 1/2)}
  \end{equation}
where $\nbar$ is the average excitation number of the atom in the harmonic
potential, and $\ld = k\sqrt{\hbar/2m_{atom}\omega_t}$ with $m_{atom}$ being
the mass of the atom and $\omega_t$ the trap frequency. The dimensionless
constant $\ld$ is called the Lamb-Dicke parameter and determines the extent to
which an atom will recoil. Since we are primarily interested in small angles
$\hat{\theta}$ and $\phi$, we expand the cosine terms in the exponent to second
order in these angles. Using this approximation, we show in
Appendix~\ref{ap:Ap_RecoilAmp} that
  \begin{eqnarray}
    \avb & = & -\sqrt{\avbmag} \frac{1-\exp{(-\ld^2(\nbar+1)\Delta^2)}}{\ld^2(\nbar +1)\Delta^2}\\ \label{eq:amp1}
    \avbmag & = & \frac{3}{8}N_s \Delta^2   \label{eq:amp2}
  \end{eqnarray}
The above expression is derived in the limit that $\Delta_i\to 0 $ and
$\Delta_o = \Delta$.  The matching condition $\alpha=\beta$ cannot be satisfied
because $\beta$ now fluctuates due to recoil.  The fidelity is maximizes when
$\alpha=-\sqrt{\avbmag}$ which indicates that $\alpha$ is set to the average
amplitude of $\beta$ and has the same phase.  Under this optimal condition, the
fidelity is given by
  \begin{equation} \label{eq:Fid}
    F = \frac{1}{2} \frac{1 + \frac{1-\exp{(-\ld^2(\nbar+1)\Delta^2)}}{\ld^2(\nbar +1)\Delta^2}}{2-\frac{1-\exp{(-\ld^2(\nbar+1)\Delta^2)}}{\ld^2(\nbar +1)\Delta^2}}
  \end{equation}
This expression allows us to calculate the fidelity of the entangled state in
the monochromatic limit as a function of the atom temperature ($\nbar$) and the
collection angle.

\begin{figure}
\centering
\includegraphics[width=5in]{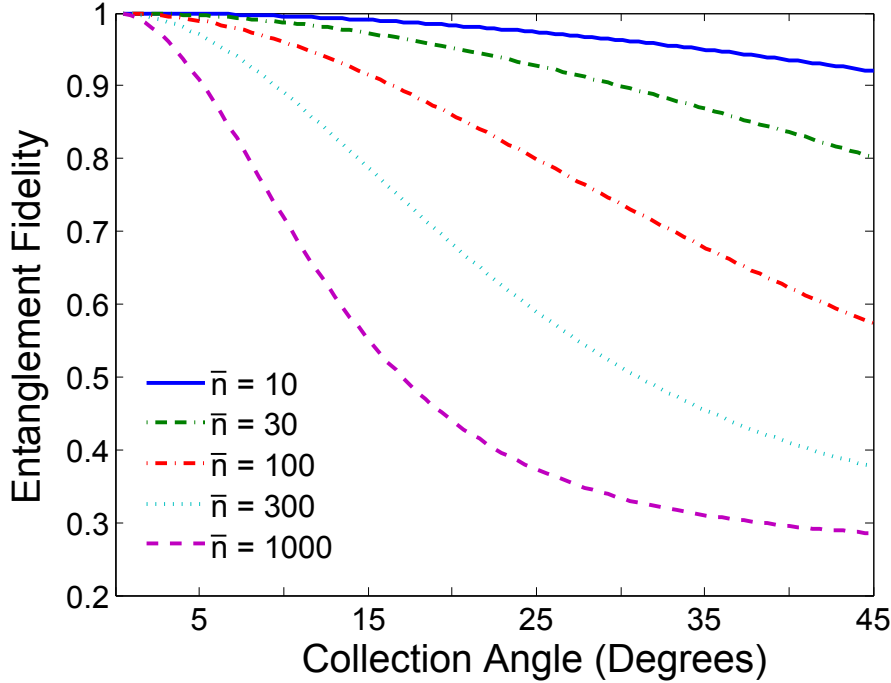}
\caption{Entanglement fidelity as a function of collection angle for the atom emission with
various atom thermal states, characterized by the average vibrational occupation
number $\nbar$. }
\label{fig:fidelity}
\end{figure}

Fig.~\ref{fig:fidelity} plots the entanglement fidelity as a function of
collected solid angle for different trap temperature. Calculations were
performed for a 1 MHz trap frequency, using the emission frequency of 935nm and
the mass of 171 atomic mass units for \Yb. For these numbers, the Lambe-Dicke
parameter is given by $\ld=0.09$, which means that the atom is in the
Lamb-Dicke regime for $\nbar<11.1$.  One can see from Fig.~\ref{fig:fidelity}
that for $\nbar=10$, the fidelity is greater than $0.9$ even for large
collection angles.  For higher trap temperatures there is a tradeoff between
the fidelity and the collection angle.

\section{Photonic Leakage of Which-Path Information} \label{sec:WhichPath}

The tradeoff between fidelity and collection angle implies that there will be a
tradeoff between fidelity and entanglement success probability, because larger
collection angle results in higher collection efficiency of scattered photons.
The entanglement success probability is given by
  \begin{equation} \label{eq:Psuccess}
    P_{success} = 1-e^{\langle |\beta|^2 \rangle /4} = 1 - e^{3 N_s \Delta^2/32}
  \end{equation}
The factor of 4 in the exponent of the above equation accounts for the fact
that each photon has a 25\% probability of triggering detector 2.

To investigate the entanglement success probability we need to determine how
large we can make $N_s$.  The expression for fidelity  in Eq.~\ref{eq:Fid} does
not depend on $N_s$ because it was derived under the assumption that the
scattered field is sufficiently weak so that it could be expanded to first
order in photon number. If we make $N_s$ too large this assumption will no
longer be valid. Thus, we need to derive a more precise expression for the
fidelity that accounts for both recoil and the probability that more than one
photon is scattered by the atom.

If the atom scatters more than one photon, this can lead to leakage of
''which-path'' information.  This information leakage can be understood from
the following simple argument.  First, we note that in general the collection
efficiency of scattered light is small.  Even with a collected solid angle of
$45^o$, which corresponds to a numerical aperture of 1, only 20\% of the
photons are collected into the fiber.  Reflection and absorption losses from
the optics will serve to degrade this collection efficiency even more.  Thus,
if the atom scatters two photons it is much more likely that only one of the
photons is collected than it is that both photons are collected.  The photon
which fails to be collected is never mixed on the beamsplitter, and therefore
retains the information that the atom caused a scattering event causing the
entanglement to decohere. Thus, in order to achieve high fidelity entanglement
the probability of scattering two photons must remain low.

To analyze multi-photon scattering we retain the amplitude of the field
scattered by the atom to all orders in photon number, but expand the field
collected into the fiber to first order in photon number. This approximation is
valid because, as noted previously, fiber collection efficiency is small in the
limit we consider so even if the atom scatters many photons the probability
that more than one of them is collected into the fiber is still small.  After
the atom scatters the input field the state of the system becomes
  \begin{equation}
    \|\psi_{ni}> = \left[ (1+\alpha\adag)(1+\beta\bdag)\|00>\|\chi> + \|11>\|0> + (1+\alpha\dag)\|01>\|0> +(1+\beta\bdag)\|10>\|\chi>\right]\|n>
  \end{equation}
In the above equation the state $\|00>\|\chi>$ denotes the atom and the QD are
both in qubit state $\|0>$, and $\chi=-\sqrt{N_s(1-3\Delta^2/8)}$ is the
coherent state amplitude of the uncollected field.  Because we consider only
small collection efficient $\chi\approx -\sqrt{N_s}$. Other states are defined
analogously, and state $\|n>$ is once again the harmonic oscillator state of
the atom center of mass. Following the same procedure as in
section~\ref{sec:Recoil}, the fidelity can be obtained by tracing over both the
field and atom center of mass motion to obtain
  \begin{equation}\label{eq:Feff}
    F = \frac{1}{2} \frac{1 + e^{-N_s/2}\frac{1-\exp (-\ld^2(\nbar+1)\Delta^2)}{\ld^2(\nbar +1)\Delta^2}}{2 - \frac{1-\exp(-\ld^2(\nbar+1)\Delta^2)}{\ld^2(\nbar +1)\Delta^2}}
  \end{equation}
The above equation gives the entanglement fidelity due to both recoil and
multi-photon scattering in the monochromatic limit.  One can see the
interference term is degraded by $e^{-N_s/2}$, which means that to achieve high
fidelity we need $N_s\ll 1$. The number of photons collected into the fiber
will be subsequently much smaller, which highlights the need for efficient
collection of photons.

\begin{figure}
  \centering
  \includegraphics[width=5in]{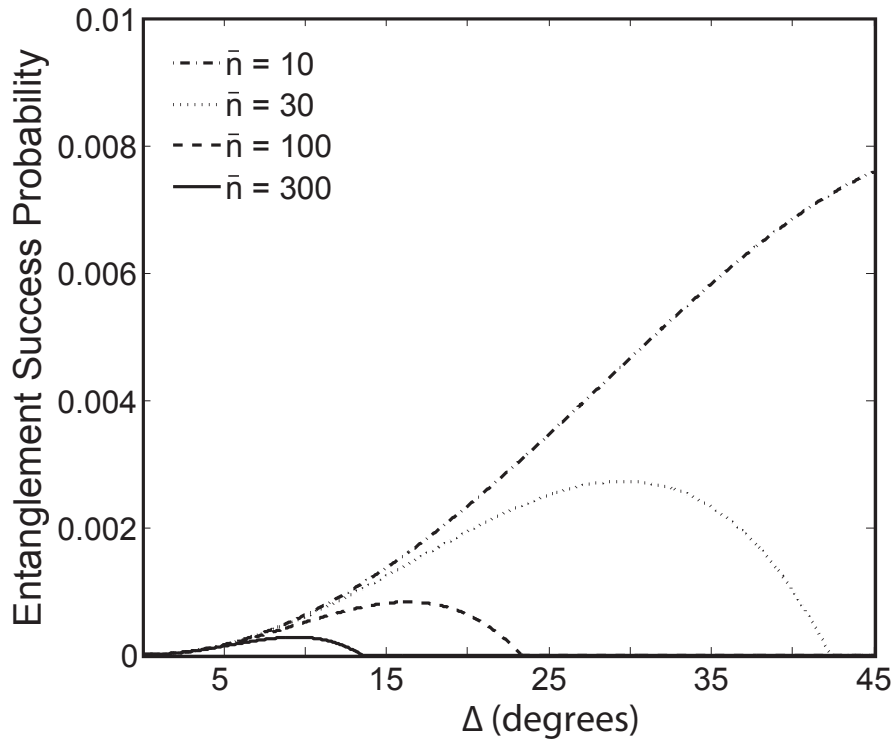}
  \caption{Success probability for creating entanglement as a function of collection angle.
  The fidelity is fixed at F=0.9.}
  \label{fig:efficiency}
  \end{figure}

For a fixed fidelity $F$, Eq.~\ref{eq:Feff} gives a relationship between the
scattered photon number $N_s$ and the collected solid angle $\Delta$.  Thus,
the entanglement success probability given in Eq.~\ref{eq:Psuccess} becomes a
function only of $\Delta$. Fig.~\ref{fig:efficiency} plot $P_{success}$ as a
function of $\Delta$ for several different trap temperatures with the same atom
trap parameters used in Sec.~\ref{sec:Recoil}, where the entanglement fidelity
is held fixed at $F=0.9$. For all trap temperatures other than $\nbar=0$, an
optimal rate exists for a specific collection angle $\Delta$.  This optimal
rate is determined by a balance between recoil and photon collection
efficiency. For $\nbar=0$ the atom is in the Lamb-Dicke regime where recoil is
not a factor, so the entanglement rate is limited only by the fraction of light
that can be collected by the optics.

\begin{figure}
  \centering
  \includegraphics[width=5in]{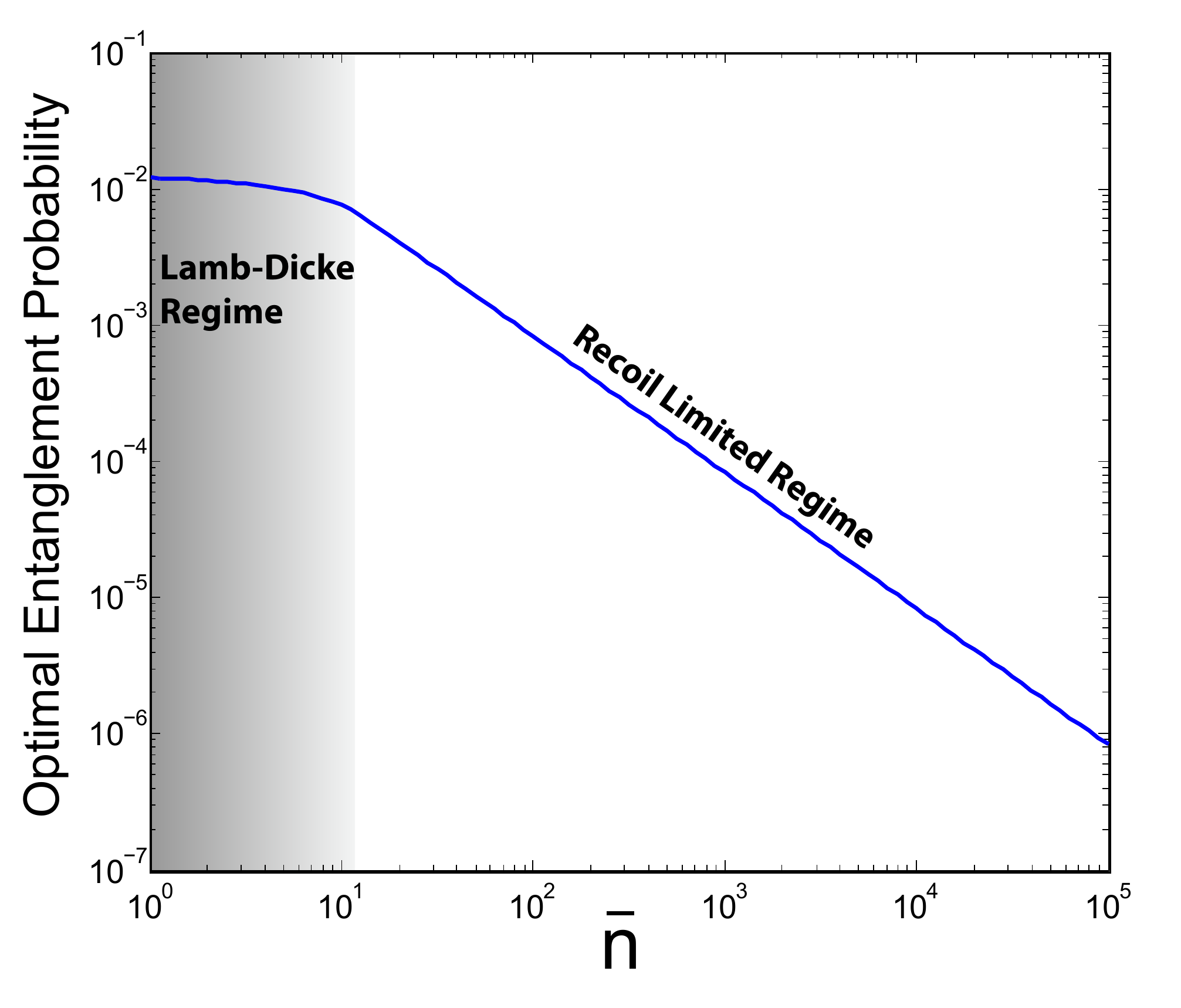}
  \caption{Optimum collection efficiency as a function of the average thermal vibration index of the trapped ion.  The
efficiency is optimized with respect to collection angle at each
point, with fidelity fixed at $F=0.9$.  The collection angle is restricted to not
exceed $45$ degrees. The shaded area represents the regime where efficiency is
limited by the collection angle of the optics.  Unshaded region represents
regime where efficiency is limited by recoil, and thus decreases as the ion gets hotter.}
  \label{fig:MaxEff}
  \end{figure}

Fig.~\ref{fig:MaxEff} plots the optimal efficiency as a function of $\nbar$.
The efficiency is optimized with respect to the collection angle for each
point, with the additional constraint that $\Delta$ cannot exceed $45^o$. For
cold states of atomic motion within the Lamb-Dicke regime $(\nbar < 11.1)$, the
efficiency is independent of $\nbar$. As the atomic motion leaves the
Lamb-Dicke regime, the collection angle must be reduced to maintain the desired
fidelity, leading to a lower success probability.  In a trap with frequency
$\omega_t/2\pi > 1$ MHz for an atomic \Yb ion, Doppler cooling is expected to
result in a mean thermal vibrational index of $\nbar < 10$, where success
probabilities can be greater than $10^{-2}$. If we use a 10 MHz experimental
repetition rate, this would result in greater than $10^5$ successful
entanglement operations per second. Additional losses to reflection from optics
and imperfect fiber coupling would serve to reduce this number.

\section{Validity of weak excitation limit} \label{sec:nonlinear}

In addition to multi-photon scattering, we must also consider the validity of
the weak excitation approximation. All of our calculations so far assumed that
the atom and cavity-QD system are driven with sufficiently weak excitation such
that $\sigma_z\to -1$, where $\sigma_z$ is the population inversion operator.
For the QD, it has been previously shown that the weak excitation limit is
valid so long as $N_{ref}/\tau_p\ll 1/\tau_{mod}$, where $N_{ref}$ is the
number of photons reflected from the cavity, $\tau_p$ is the input pulse
duration, and $\tau_{mod}$ is the modified spontaneous emission lifetime of the
QD~\cite{SridharanWaks2008}.  For any $N_{ref}$ one can in principle satisfy
weak excitation by make $\tau_p$ sufficiently long. It should also be noted
that from the matching condition $\alpha = \beta$ and the condition that
$N_s\ll 1$ (due to multi-photon scattering) we know that $N_{ref}\ll 1$.  In
addition, for the input pulse spectrum to fit within the high-reflection window
given in Eq.~\ref{eq:r}, $\tau_p > \tau_{mod}$~\cite{WaksVuckovicPRL2006}. When
these two conditions are combined they automatically guarantee that weak
excitation is satisfied for the QD. Thus, for the QD weak excitation does not
impose any additional constraints on the entanglement success probability and
can be generally satisfied by using sufficiently long input pulses.

We now derive a similar result for the atom.  We use Eq.~\ref{eq:sigmamp} to
show that
  \begin{equation}\label{eq:weak1}
    \langle \sigma_+\sigma_-\rangle = \frac{|L(\omega_0)|^2}{\gamma^2} |\Omega(\r,t)|^2
  \end{equation}
where $\Omega(\r,t) = d E(\r,t)/\hbar$.  The weak excitation approximation is
valid so long as $\langle \sigma_+\sigma_-\rangle\ll 1$.  This condition can be
recast into a more recognizable form. The easiest way to do this is to assume
that the input optical pulse is a square pulse of duration $\tau$ starting at
$t=0$, with electric field amplitude $E$. Thus
  \begin{equation}
    \int_0^\infty |E(\r,t)|^2 dt = |E|^2\tau = \frac{n\hbar\omega_0}{c\eo}
  \end{equation}
where the last equality comes from Poynting's theorem.  Eq. ~\ref{eq:weak1}
becomes
  \begin{equation} \label{eq:nonlin}
    \frac{n\sigma_0|L(\omega_0)|^2\left(\frac{\gamma_r}{\gamma}\right)^2}{\tau} = \frac{N_s}{\tau} \ll \gamma
  \end{equation}
We attain a condition for the atom which states that the rate of scattered
photons should be small compared to the atom decay rate. Although this
condition was attained using the assumption of a square pulse, we expect this
relation to hold for most pulse shapes.

As was the case for the QD, the limitations imposed by weak excitation restrict
only the rate of emitted photons, not on total photon number. In this way, weak
excitation provides a weaker restriction than multi-photon scattering, and can
in general be well satisfied by picking sufficiently long pulses. In a
practical experiment the clock cycle for generating entanglement will almost
always be long compared to the atom decay rate.  This ensures that all
transients have decayed between consecutive cycles of the experiment.  If we
combine this requirement with the restriction placed by multi-photon scattering
that $N_s\ll 1$, then Eq.~\ref{eq:nonlin} is automatically satisfied. Thus, at
the point where nonlinearities become important the system will already have
decohered due to multi-photon scattering.

\section{Conclusion}

We conclude that it appears feasible to entangle a QD and an atom by weakly
scattering light from each system and interfering these fields to produce an
appropriate single-photon detection event that heralds the entanglement.
Differential dispersion, atom recoil, and the multi-photon scattering can all
be managed by properly selecting the input pulse duration, collected solid
angle, and input pump power.  It is noted that to implement the proposed
protocol it is necessary to overcome difficult experimental challenges.  For
example, the protocol requires phase-locking of all optical pulses for qubit
rotation of both atom and QD, which will require pulling all optical pulses
from a common laser source or using multiple phase-locked lasers, adding to the
experimental difficulty.  In addition, decay of the atomic \Yb to other
transitions will require periodic re-pumping into the \Dtwo manifold.
Nevertheless, the work we present suggests that entanglement between an atomic
and semiconductor system is within the reach of presently available
technological capabilities.

\acknowledgements

This work is supported by the Intelligence Advanced Research Projects Activity
(IARPA) under Army Research Office contract, the NSF Physics at the Information
Frontier Program, the NSF Physics Frontier Center at JQI, and an Army Research
Office Young Investigator Award.

\appendix
\section{Calculation of Total Scattered Photon Number} \label{app:NumScat}

The average number of scattered photons can be calculated by $N=\sum_\k \langle
\bkdag\bk\rangle = \sum_k |\beta_\k|^2$.  Plugging Eq.~\ref{eq:ScatAmp} into
this expression we obtain
  \begin{eqnarray}
    N & = & \sum_\k \pi^2 \frac{g_\k^2 |L(\omega_0)|^2}{\gamma^2} \Omega(\omega_\k) e^{i(\k_\omega - \k)\cdot\r} \\
    & \to & \frac{V}{2\pi c}\int_0^{2\pi}d\phi\int_0^{\pi}d\theta\int d\omega \omega^2 \sin\theta \pi^2 \frac{g_\k^2 |L(\omega_0)|^2}{\gamma^2} \Omega(\omega_\k) e^{i(\k_\omega - \k)\cdot\r} \\
    & = & \frac{|L(\omega_0)|^2}{\gamma^2} \frac{d^4\omega_0^3}{6\hbar^3\eo c^3} \int_{-\infty}^\infty d\omega |E(\omega)|^2 \label{eq:eqscat}
  \end{eqnarray}
The amount of energy in the pump beam can be determines using Poynting's
theorem:
  \begin{eqnarray*}
    W & = & \int_0^{\infty} dt \oint \frac{\E\times\B}{\uo} \cdot d\a \\
    & = & \frac{A}{c\uo} \int_0^\infty E^2(t) dt
  \end{eqnarray*}
where $A$ is the cross sectional area of the pump beam. Using the definition
  \begin{equation}
    E(t) = \int d\omega E(\omega)e^{-i\omega t}
  \end{equation}
we attain
  \begin{equation}
    W=A\pi c\eo\int d\omega |E(\omega)|^2 = N_i \hbar\omega_0
  \end{equation}
where $N_i$ is the total number of photons in the pump and $\omega_0$ is the
center frequency of the quasi-monochromatic pump beam.  Defining the photon
density $n_i=N_i/A$ we then have
  \begin{equation}
    \int d\omega |E(\omega)|^2 =  \frac{\hbar \omega_0 n_i}{\pi c \eo}
  \end{equation}
Plugging this expression back into Eq.~\ref{eq:eqscat}, and using
  \begin{equation} \label{eq:gamma}
    \gamma_r = \frac{\omega_0^3 d^2}{6\pi\eo\hbar c^3}
  \end{equation}
leads directly to Eq.~\ref{eq:NumScat}.

\section{Field amplitude collected into optical fiber} \label{ap:FiberAmp}

For each frequency $\omega$ we define a fiber mode $\bw$.  The collection lens
and fiber are linear optical components, which means that the fiber mode is
related to the free space modes by the linear transformation
  \begin{equation}
    \bw = \sum_{k=\omega/c} s_\k \bk
  \end{equation}
The sum is carried out over all free space modes that have the same energy as
the fiber mode due to linearity.  Unitarity requires that
  \begin{equation} \label{eq:Unitary}
    \sum_{k=\omega/c} |s_\k|^2 = 1
  \end{equation}
We define $\Delta\theta$ as the angle between the $\k$ and the input field
propagation direction which is assumed to be along the $x$ axis.  We adopt a
simplified model that the collection optics collect all k-vectors satisfying
$\Delta_i < \Delta\theta < \Delta_o$.  Thus, for k-vectors satisfying this
condition $s_\k=C$, where $C$ is a constant, while $s_\k = 0$ for all other
k-vectors.

The constant $C$ must be determined from the condition Eq.~\ref{eq:Unitary},
which results in
  \begin{equation}
    \sum_{k=\omega/c} |C|^2 = 1
  \end{equation}
To evaluate the above sum we make the additional assumption that the lens
collects $\k$ vectors propagating very close to the x-axis.  In this
\textit{paraxial wave} limit, the sum can be converted to an integral as
  \begin{equation}
    \sum_{k=\omega/c} |C|^2 = |C|^2 \frac{L_y L_z}{(2\pi c)^2} \int d\theta \int d\phi \omega^2 \sin\theta
  \end{equation}
where $L_y$ and $L_z$ are the length of the bounding box from the quantization
of the free space modes in the $y$ and $z$ directions, and angles $\theta$ and
$\phi$ are defined in Fig.~\ref{fig:collection}. From the above equation we
attain
  \begin{equation}
    C = \frac{2\pi c}{\omega} \frac{1}{\sqrt{L_y L_z}} \frac{1}{\sqrt{A}}
  \end{equation}
where $A=2\pi(\cos\Delta_i - \cos\Delta_o)\approx \pi(\Delta_o^2 -
\Delta_i^2)$.  Using this expression we then have
  \begin{eqnarray*}
    \beta_\omega & = & \langle\bw\rangle \\
    & = & \frac{2\pi c}{\omega} \frac{1}{\sqrt{L_y L_z}} \frac{1}{\sqrt{A}} \sum_{[\theta,\phi]\in [\Dx,\Dy]} \langle \bk \rangle \\
  \end{eqnarray*}
where $\bk$ is given in Eq.~\ref{eq:bksolved}. We again turn the sum into an
integral and perform some algebraic manipulation to attain
  \begin{equation}
    \beta_\omega = -\frac{ d^2 \omega^{3/2} \Lor[\delta_a,\gamma_a]\Omega(\omega) e^{-i\omega t}}{2c \gamma_a \sqrt{2\eo\hbar L_x  A}}  \int d\theta \int d\phi e^{i(\k_{\omega} - \k)\cdot\r}
  \end{equation}
where angles $\theta$ and $\phi$ are illustrated in Fig.~\ref{fig:collection}.
We assume that the input pulse has a narrow bandwidth centered around
$\omega_0$ so that we may make the substitutions $\omega^{3/2}\approx
\omega_0^{3/2}$ and $\exp (\k_\omega - \k)\approx \exp [k_0 (\xhat - \khat)]$.
We do not, however, make this approximation for $\Lor[\delta_a,\gamma_a]$ which
is a rapidly varying function of $\omega$ near resonance.  With these
approximations we attain the result stated in Eq.~\ref{eq:beta} and
Eq.~\ref{eq:beta0}.

\section{Calculation of average scattering amplitude due to atomic
recoil}\label{ap:Ap_RecoilAmp}

We assume that the collection optics are frequency independent over the
bandwidth of the collected signal. We assume quasi-monochromatic input so that
$\Lor[\delta_a,\gamma_a]= \gamma_a/[\gamma_a-i(\omega_a - \omega_0)]$. We can
construct a fiber mode of the form
  \begin{equation} \label{eq:mode}
    \b = \sum_\omega \chi E^*(\omega)e^{i\omega t} \bw
  \end{equation}
where $\chi$ is a normalization constant.  This is the only mode that the
collected field will couple to.  To understand why, we first note that we can
construct a complete basis using the above mode along with a set of other
orthogonal modes that can be calculated using Schmidt decomposition.  We can
then calculate the field amplitude using
  \begin{equation} \label{eq:Betacalc}
  \langle \beta \rangle = \langle \b \rangle = \sum_\omega \chi E^*(\omega)e^{i\omega t} \langle \beta_\omega \rangle
  \end{equation}
Since $\beta_\omega$ is proportional to the complex conjugate of the expansion
coefficient $\chi E^*(\omega)e^{i\omega t}$, it will have a maximum overlap
with the mode in Eq.~\ref{eq:mode}, and will be orthogonal to all other modes,
ensuring the mode in Eq.~\ref{eq:mode} is the only one that contains photons.

Unitarity determines the value of $\chi$ from
  \begin{equation}
    \sum_\omega |\chi|^2 |E(\omega)|^2 = 1
  \end{equation}
which leads to
  \begin{equation}
    \chi = \frac{\pi c \sqrt{2\eo}}{\sqrt{L_x \hbar \omega_0}} \frac{1}{\sqrt{n}}
  \end{equation}
Plugging the above expression into Eq.~\ref{eq:Betacalc} and turning the sum
into an integral we obtain
  \begin{equation}
    \langle \beta \rangle = B\int d\theta \int d\phi \langle e^{i(\k_{\omega_0} - \k)\cdot\r} \rangle
  \end{equation}
where
  \begin{equation}
    B = \frac{-d^2\omega_0^2}{4\pi c^2 \eo \hbar}\sqrt{\frac{n}{A}}\frac{\Lor[\delta_a,\gamma_a]}{\gamma_a}
  \end{equation}
We now use relation Eq.~\ref{eq:ThermAvg} to write
  \begin{equation}
    \langle \beta \rangle = B \int d\theta \int d\phi e^{-2\ld^2 (1-\cos\hat{\theta}\cos\phi) (\nbar + 1/2)}
  \end{equation}
Because we are working in the paraxial limit we can expand the exponent to
second order in $\hat{\theta}$ and $\phi$.  Defining $\Delta =
\sqrt{\hat{\theta}^2 + \phi^2}$ we have
  \begin{equation}
    \int d\theta \int d\phi e^{-2\ld^2 (1-\cos\hat{\theta}\cos\phi) (\nbar + 1/2)} = 2\pi\int_{\Delta_i}^{\Delta_o} d\Delta \Delta e^{-\ld^2 \Delta^2} = \frac{e^{-\ld^2 \Delta_o^2} - e^{-\ld^2 \Delta_i^2}}{\ld^2}
  \end{equation}
Taking the limit $\Delta_i\to 0 $ and plugging into the above equation we
attain the expression in Eq.~\ref{eq:avb}.

Similarly, we can write
  \begin{equation}
    \langle |\beta|^2\rangle = |B|^2 \int d\theta \int d\phi \int d\theta' \int d\phi' \langle e^{-(\k - \k')\cdot\r}\rangle
  \end{equation}
It is straightforward to show that the lowest order contribution to the
exponent is fourth order in $\theta$, $\theta'$, $\phi$, and $\phi'$.  Since we
are only expanding to second order in these variables, we have
  \begin{equation}
    \langle |\beta|^2\rangle = |B|^2 \int d\theta \int d\phi \int d\theta' \int d\phi' = |B|^2 A^2
  \end{equation}
which leads directly to Eq.~\ref{eq:avbmag}.

%\bibliography{IonQD3}

\begin{thebibliography}{34}
\expandafter\ifx\csname natexlab\endcsname\relax\def\natexlab#1{#1}\fi
\expandafter\ifx\csname bibnamefont\endcsname\relax
  \def\bibnamefont#1{#1}\fi
\expandafter\ifx\csname bibfnamefont\endcsname\relax
  \def\bibfnamefont#1{#1}\fi
\expandafter\ifx\csname citenamefont\endcsname\relax
  \def\citenamefont#1{#1}\fi
\expandafter\ifx\csname url\endcsname\relax
  \def\url#1{\texttt{#1}}\fi
\expandafter\ifx\csname urlprefix\endcsname\relax\def\urlprefix{URL }\fi
\providecommand{\bibinfo}[2]{#2} \providecommand{\eprint}[2][]{\url{#2}}

\bibitem[{\citenamefont{Einstein et~al.}(1935)\citenamefont{Einstein, Podolsky,
  and Rosen}}]{EPR}
\bibinfo{author}{\bibfnamefont{A.}~\bibnamefont{Einstein}},
  \bibinfo{author}{\bibfnamefont{B.}~\bibnamefont{Podolsky}}, \bibnamefont{and}
  \bibinfo{author}{\bibfnamefont{N.}~\bibnamefont{Rosen}},
  \bibinfo{journal}{Phys. Rev.} \textbf{\bibinfo{volume}{47}},
  \bibinfo{pages}{777} (\bibinfo{year}{1935}).

\bibitem[{\citenamefont{Schroedinger}(1936)}]{Schrodinger}
    \bibinfo{author}{\bibfnamefont{E.}~\bibnamefont{Schroedinger}},
  \bibinfo{journal}{Proc. Cambridge Phil. Soc.} \textbf{\bibinfo{volume}{31}},
  \bibinfo{pages}{555} (\bibinfo{year}{1936}).

\bibitem[{\citenamefont{Nielsen and Chuang}(2000)}]{nielsen}
    \bibinfo{author}{\bibfnamefont{M.~A.} \bibnamefont{Nielsen}}
    \bibnamefont{and}
  \bibinfo{author}{\bibfnamefont{I.~L.} \bibnamefont{Chuang}},
  \emph{\bibinfo{title}{Quantum Computation and Quantum Information}}
  (\bibinfo{publisher}{Cambridge University Press},
  \bibinfo{address}{Cambridge, UK}, \bibinfo{year}{2000}).

\bibitem[{\citenamefont{Monroe and Lukin}(August, 2008)}]{PW08}
    \bibinfo{author}{\bibfnamefont{C.}~\bibnamefont{Monroe}} \bibnamefont{and}
  \bibinfo{author}{\bibfnamefont{M.}~\bibnamefont{Lukin}},
  \bibinfo{journal}{Physics World} pp. \bibinfo{pages}{32--39}
  (\bibinfo{year}{August, 2008}).

\bibitem[{\citenamefont{Moehring et~al.}(2004)\citenamefont{Moehring, Madsen,
  Blinov, and Monroe}}]{Moehring2004}
\bibinfo{author}{\bibfnamefont{D.~L.} \bibnamefont{Moehring}},
  \bibinfo{author}{\bibfnamefont{M.~J.} \bibnamefont{Madsen}},
  \bibinfo{author}{\bibfnamefont{B.~B.} \bibnamefont{Blinov}},
  \bibnamefont{and} \bibinfo{author}{\bibfnamefont{C.}~\bibnamefont{Monroe}},
  \bibinfo{journal}{Physical Review Letters} \textbf{\bibinfo{volume}{93}},
  \bibinfo{pages}{090410} (\bibinfo{year}{2004}), \bibinfo{note}{copyright (C)
  2009 The American Physical Society Please report any problems to
  prola@aps.org PRL}.

\bibitem[{\citenamefont{Sherson et~al.}(2006)\citenamefont{Sherson, Julsgaard,
  and Polzik}}]{Sherson06}
\bibinfo{author}{\bibfnamefont{J.}~\bibnamefont{Sherson}},
  \bibinfo{author}{\bibfnamefont{B.}~\bibnamefont{Julsgaard}},
  \bibnamefont{and} \bibinfo{author}{\bibfnamefont{E.~S.}
  \bibnamefont{Polzik}}, \bibinfo{journal}{Adv. At. Mol. Opt. Phys.}
  \textbf{\bibinfo{volume}{54}}, \bibinfo{pages}{82} (\bibinfo{year}{2006}).

\bibitem[{\citenamefont{Schmidt et~al.}(2005)\citenamefont{Schmidt, Rosenband,
  Langer, Itano, Bergquist, and Wineland}}]{Schmidt05}
\bibinfo{author}{\bibfnamefont{P.~O.} \bibnamefont{Schmidt}},
  \bibinfo{author}{\bibfnamefont{T.}~\bibnamefont{Rosenband}},
  \bibinfo{author}{\bibfnamefont{C.}~\bibnamefont{Langer}},
  \bibinfo{author}{\bibfnamefont{W.~M.} \bibnamefont{Itano}},
  \bibinfo{author}{\bibfnamefont{J.~C.} \bibnamefont{Bergquist}},
  \bibnamefont{and} \bibinfo{author}{\bibfnamefont{D.~J.}
  \bibnamefont{Wineland}}, \bibinfo{journal}{Science}
  \textbf{\bibinfo{volume}{309}}, \bibinfo{pages}{749} (\bibinfo{year}{2005}).

\bibitem[{\citenamefont{Duan et~al.}(2001)\citenamefont{Duan, Lukin, Cirac, and
  Zoller}}]{DLCZ01}
\bibinfo{author}{\bibfnamefont{L.~M.} \bibnamefont{Duan}},
  \bibinfo{author}{\bibfnamefont{M.~D.} \bibnamefont{Lukin}},
  \bibinfo{author}{\bibfnamefont{J.~I.} \bibnamefont{Cirac}}, \bibnamefont{and}
  \bibinfo{author}{\bibfnamefont{P.}~\bibnamefont{Zoller}},
  \bibinfo{journal}{Nature} \textbf{\bibinfo{volume}{414}},
  \bibinfo{pages}{413} (\bibinfo{year}{2001}), \bibinfo{note}{0028-0836
  10.1038/35106500 10.1038/35106500}.

\bibitem[{\citenamefont{Childress et~al.}(2005)\citenamefont{Childress, Taylor,
  Sorensen, and Lukin}}]{Childress06}
\bibinfo{author}{\bibfnamefont{L.}~\bibnamefont{Childress}},
  \bibinfo{author}{\bibfnamefont{J.~M.} \bibnamefont{Taylor}},
  \bibinfo{author}{\bibfnamefont{A.~S.} \bibnamefont{Sorensen}},
  \bibnamefont{and} \bibinfo{author}{\bibfnamefont{M.~D.} \bibnamefont{Lukin}},
  \bibinfo{journal}{Phys. Rev. A} \textbf{\bibinfo{volume}{72}},
  \bibinfo{eid}{052330} (pages~\bibinfo{numpages}{16}) (\bibinfo{year}{2005}).

\bibitem[{\citenamefont{Waks and Vuckovic}(2006)}]{WaksVuckovicPRL2006}
    \bibinfo{author}{\bibfnamefont{E.}~\bibnamefont{Waks}} \bibnamefont{and}
  \bibinfo{author}{\bibfnamefont{J.}~\bibnamefont{Vuckovic}},
  \bibinfo{journal}{Physical Review Letters} \textbf{\bibinfo{volume}{96}},
  \bibinfo{eid}{153601} (pages~\bibinfo{numpages}{4}) (\bibinfo{year}{2006}),
  \urlprefix\url{http://link.aps.org/abstract/PRL/v96/e153601}.

\bibitem[{\citenamefont{Sridharan and Waks}(2008)}]{SridharanWaks2008}
    \bibinfo{author}{\bibfnamefont{D.}~\bibnamefont{Sridharan}}
    \bibnamefont{and}
  \bibinfo{author}{\bibfnamefont{E.}~\bibnamefont{Waks}},
  \bibinfo{journal}{Physical Review A (Atomic, Molecular, and Optical Physics)}
  \textbf{\bibinfo{volume}{78}}, \bibinfo{eid}{052321}
  (pages~\bibinfo{numpages}{10}) (\bibinfo{year}{2008}),
  \urlprefix\url{http://link.aps.org/abstract/PRA/v78/e052321}.

\bibitem[{\citenamefont{Moehring et~al.}(2007)\citenamefont{Moehring, Maunz,
  Olmschenk, Younge, Matsukevich, Duan, and Monroe}}]{Moehring07}
\bibinfo{author}{\bibfnamefont{D.~L.} \bibnamefont{Moehring}},
  \bibinfo{author}{\bibfnamefont{P.}~\bibnamefont{Maunz}},
  \bibinfo{author}{\bibfnamefont{S.}~\bibnamefont{Olmschenk}},
  \bibinfo{author}{\bibfnamefont{K.~C.} \bibnamefont{Younge}},
  \bibinfo{author}{\bibfnamefont{D.~N.} \bibnamefont{Matsukevich}},
  \bibinfo{author}{\bibfnamefont{L.-M.} \bibnamefont{Duan}}, \bibnamefont{and}
  \bibinfo{author}{\bibfnamefont{C.}~\bibnamefont{Monroe}},
  \bibinfo{journal}{Nature} \textbf{\bibinfo{volume}{449}}, \bibinfo{pages}{68}
  (\bibinfo{year}{2007}).

\bibitem[{\citenamefont{Chou et~al.}(2005)\citenamefont{Chou, de~Riedmatten,
  Felinto, Polyakov, van Enk, and Kimble}}]{ChouRiedmatten2005}
\bibinfo{author}{\bibfnamefont{C.~W.} \bibnamefont{Chou}},
  \bibinfo{author}{\bibfnamefont{H.}~\bibnamefont{de~Riedmatten}},
  \bibinfo{author}{\bibfnamefont{D.}~\bibnamefont{Felinto}},
  \bibinfo{author}{\bibfnamefont{S.~V.} \bibnamefont{Polyakov}},
  \bibinfo{author}{\bibfnamefont{S.~J.} \bibnamefont{van Enk}},
  \bibnamefont{and} \bibinfo{author}{\bibfnamefont{H.~J.}
  \bibnamefont{Kimble}}, \bibinfo{journal}{Nature}
  \textbf{\bibinfo{volume}{438}}, \bibinfo{pages}{828} (\bibinfo{year}{2005}),
  \bibinfo{note}{10.1038/nature04353}.

\bibitem[{\citenamefont{Chou et~al.}(2007)\citenamefont{Chou, Laurat, Deng,
  Choi, de~Riedmatten, Felinto, and Kimble}}]{ChouJulien2007}
\bibinfo{author}{\bibfnamefont{C.-W.} \bibnamefont{Chou}},
  \bibinfo{author}{\bibfnamefont{J.}~\bibnamefont{Laurat}},
  \bibinfo{author}{\bibfnamefont{H.}~\bibnamefont{Deng}},
  \bibinfo{author}{\bibfnamefont{K.~S.} \bibnamefont{Choi}},
  \bibinfo{author}{\bibfnamefont{H.}~\bibnamefont{de~Riedmatten}},
  \bibinfo{author}{\bibfnamefont{D.}~\bibnamefont{Felinto}}, \bibnamefont{and}
  \bibinfo{author}{\bibfnamefont{H.~J.} \bibnamefont{Kimble}},
  \bibinfo{journal}{Science} \textbf{\bibinfo{volume}{316}},
  \bibinfo{pages}{1316} (\bibinfo{year}{2007}).

\bibitem[{\citenamefont{Choi et~al.}(2008)\citenamefont{Choi, Deng, Laurat, and
  Kimble}}]{ChoiDeng2008}
\bibinfo{author}{\bibfnamefont{K.~S.} \bibnamefont{Choi}},
  \bibinfo{author}{\bibfnamefont{H.}~\bibnamefont{Deng}},
  \bibinfo{author}{\bibfnamefont{J.}~\bibnamefont{Laurat}}, \bibnamefont{and}
  \bibinfo{author}{\bibfnamefont{H.~J.} \bibnamefont{Kimble}},
  \bibinfo{journal}{Nature} \textbf{\bibinfo{volume}{452}}, \bibinfo{pages}{67}
  (\bibinfo{year}{2008}), \bibinfo{note}{10.1038/nature06670}.

\bibitem[{\citenamefont{Xu et~al.}(2008{\natexlab{a}})\citenamefont{Xu, Sun,
  Berman, Steel, Bracker, Gammon, and Sham}}]{XuSun2008}
\bibinfo{author}{\bibfnamefont{X.}~\bibnamefont{Xu}},
  \bibinfo{author}{\bibfnamefont{B.}~\bibnamefont{Sun}},
  \bibinfo{author}{\bibfnamefont{P.~R.} \bibnamefont{Berman}},
  \bibinfo{author}{\bibfnamefont{D.~G.} \bibnamefont{Steel}},
  \bibinfo{author}{\bibfnamefont{A.~S.} \bibnamefont{Bracker}},
  \bibinfo{author}{\bibfnamefont{D.}~\bibnamefont{Gammon}}, \bibnamefont{and}
  \bibinfo{author}{\bibfnamefont{L.~J.} \bibnamefont{Sham}},
  \bibinfo{journal}{Nat Phys} \textbf{\bibinfo{volume}{4}},
  \bibinfo{pages}{692} (\bibinfo{year}{2008}{\natexlab{a}}),
  \bibinfo{note}{1745-2473 10.1038/nphys1054 10.1038/nphys1054}.

\bibitem[{\citenamefont{Press et~al.}(2008{\natexlab{a}})\citenamefont{Press,
  Ladd, Zhang, and Yamamoto}}]{PressLadd2008}
\bibinfo{author}{\bibfnamefont{D.}~\bibnamefont{Press}},
  \bibinfo{author}{\bibfnamefont{T.~D.} \bibnamefont{Ladd}},
  \bibinfo{author}{\bibfnamefont{B.}~\bibnamefont{Zhang}}, \bibnamefont{and}
  \bibinfo{author}{\bibfnamefont{Y.}~\bibnamefont{Yamamoto}},
  \bibinfo{journal}{Nature} \textbf{\bibinfo{volume}{456}},
  \bibinfo{pages}{218} (\bibinfo{year}{2008}{\natexlab{a}}),
  \bibinfo{note}{0028-0836 10.1038/nature07530 10.1038/nature07530}.

\bibitem[{\citenamefont{Rakher et~al.}(2009)\citenamefont{Rakher, Stoltz,
  Coldren, Petroff, and Bouwmeester}}]{rakher:097403}
\bibinfo{author}{\bibfnamefont{M.~T.} \bibnamefont{Rakher}},
  \bibinfo{author}{\bibfnamefont{N.~G.} \bibnamefont{Stoltz}},
  \bibinfo{author}{\bibfnamefont{L.~A.} \bibnamefont{Coldren}},
  \bibinfo{author}{\bibfnamefont{P.~M.} \bibnamefont{Petroff}},
  \bibnamefont{and}
  \bibinfo{author}{\bibfnamefont{D.}~\bibnamefont{Bouwmeester}},
  \bibinfo{journal}{Physical Review Letters} \textbf{\bibinfo{volume}{102}},
  \bibinfo{eid}{097403} (pages~\bibinfo{numpages}{4}) (\bibinfo{year}{2009}),
  \urlprefix\url{http://link.aps.org/abstract/PRL/v102/e097403}.

\bibitem[{\citenamefont{Wineland and Blatt}(2008)}]{WinelandBlatt08}
    \bibinfo{author}{\bibfnamefont{D.}~\bibnamefont{Wineland}}
    \bibnamefont{and}
  \bibinfo{author}{\bibfnamefont{R.}~\bibnamefont{Blatt}},
  \bibinfo{journal}{Nature} \textbf{\bibinfo{volume}{453}},
  \bibinfo{pages}{1008} (\bibinfo{year}{2008}).

\bibitem[{\citenamefont{Olmschenk et~al.}(2007)\citenamefont{Olmschenk, Younge,
  Moehring, Matsukevich, Maunz, and Monroe}}]{Yb-qubit}
\bibinfo{author}{\bibfnamefont{S.}~\bibnamefont{Olmschenk}},
  \bibinfo{author}{\bibfnamefont{K.~C.} \bibnamefont{Younge}},
  \bibinfo{author}{\bibfnamefont{D.~L.} \bibnamefont{Moehring}},
  \bibinfo{author}{\bibfnamefont{D.~N.} \bibnamefont{Matsukevich}},
  \bibinfo{author}{\bibfnamefont{P.}~\bibnamefont{Maunz}}, \bibnamefont{and}
  \bibinfo{author}{\bibfnamefont{C.}~\bibnamefont{Monroe}},
  \bibinfo{journal}{Physical Review A (Atomic, Molecular, and Optical Physics)}
  \textbf{\bibinfo{volume}{76}}, \bibinfo{eid}{052314}
  (pages~\bibinfo{numpages}{9}) (\bibinfo{year}{2007}).

\bibitem[{\citenamefont{Fu et~al.}(2008)\citenamefont{Fu, Clark, Santori,
  Stanley, Holland, and Yamamoto}}]{FuClark08}
\bibinfo{author}{\bibfnamefont{K.-M.~C.} \bibnamefont{Fu}},
  \bibinfo{author}{\bibfnamefont{S.~M.} \bibnamefont{Clark}},
  \bibinfo{author}{\bibfnamefont{C.}~\bibnamefont{Santori}},
  \bibinfo{author}{\bibfnamefont{C.~R.} \bibnamefont{Stanley}},
  \bibinfo{author}{\bibfnamefont{M.~C.} \bibnamefont{Holland}},
  \bibnamefont{and} \bibinfo{author}{\bibfnamefont{Y.}~\bibnamefont{Yamamoto}},
  \bibinfo{journal}{Nat Phys} \textbf{\bibinfo{volume}{4}},
  \bibinfo{pages}{780} (\bibinfo{year}{2008}), \bibinfo{note}{1745-2473
  10.1038/nphys1052 10.1038/nphys1052}.

\bibitem[{\citenamefont{Press et~al.}(2008{\natexlab{b}})\citenamefont{Press,
  Ladd, Zhang, and Yamamoto}}]{PressLadd08}
\bibinfo{author}{\bibfnamefont{D.}~\bibnamefont{Press}},
  \bibinfo{author}{\bibfnamefont{T.~D.} \bibnamefont{Ladd}},
  \bibinfo{author}{\bibfnamefont{B.}~\bibnamefont{Zhang}}, \bibnamefont{and}
  \bibinfo{author}{\bibfnamefont{Y.}~\bibnamefont{Yamamoto}},
  \bibinfo{journal}{Nature} \textbf{\bibinfo{volume}{456}},
  \bibinfo{pages}{218} (\bibinfo{year}{2008}{\natexlab{b}}),
  \bibinfo{note}{0028-0836 10.1038/nature07530 10.1038/nature07530}.

\bibitem[{\citenamefont{Xu et~al.}(2008{\natexlab{b}})\citenamefont{Xu, Sun,
  Berman, Steel, Bracker, Gammon, and Sham}}]{XuSun08}
\bibinfo{author}{\bibfnamefont{X.}~\bibnamefont{Xu}},
  \bibinfo{author}{\bibfnamefont{B.}~\bibnamefont{Sun}},
  \bibinfo{author}{\bibfnamefont{P.~R.} \bibnamefont{Berman}},
  \bibinfo{author}{\bibfnamefont{D.~G.} \bibnamefont{Steel}},
  \bibinfo{author}{\bibfnamefont{A.~S.} \bibnamefont{Bracker}},
  \bibinfo{author}{\bibfnamefont{D.}~\bibnamefont{Gammon}}, \bibnamefont{and}
  \bibinfo{author}{\bibfnamefont{L.~J.} \bibnamefont{Sham}},
  \bibinfo{journal}{Nat Phys} \textbf{\bibinfo{volume}{4}},
  \bibinfo{pages}{692} (\bibinfo{year}{2008}{\natexlab{b}}),
  \bibinfo{note}{1745-2473 10.1038/nphys1054 10.1038/nphys1054}.

\bibitem[{\citenamefont{Itano et~al.}(1998)\citenamefont{Itano, Bergquist,
  Bollinger, Wineland, Eichmann, and Raizen}}]{itano:1998}
\bibinfo{author}{\bibfnamefont{W.}~\bibnamefont{Itano}},
  \bibinfo{author}{\bibfnamefont{J.}~\bibnamefont{Bergquist}},
  \bibinfo{author}{\bibfnamefont{J.}~\bibnamefont{Bollinger}},
  \bibinfo{author}{\bibfnamefont{D.}~\bibnamefont{Wineland}},
  \bibinfo{author}{\bibfnamefont{U.}~\bibnamefont{Eichmann}}, \bibnamefont{and}
  \bibinfo{author}{\bibfnamefont{M.}~\bibnamefont{Raizen}},
  \bibinfo{journal}{Phys. Rev. A} \textbf{\bibinfo{volume}{57}},
  \bibinfo{pages}{4176} (\bibinfo{year}{1998}).

\bibitem[{\citenamefont{Eichmann et~al.}(1993)\citenamefont{Eichmann,
  Bergquist, Bollinger, Gilligan, Itano, Wineland, and Raizen}}]{eichmann:1993}
\bibinfo{author}{\bibfnamefont{U.}~\bibnamefont{Eichmann}},
  \bibinfo{author}{\bibfnamefont{J.~C.} \bibnamefont{Bergquist}},
  \bibinfo{author}{\bibfnamefont{J.~J.} \bibnamefont{Bollinger}},
  \bibinfo{author}{\bibfnamefont{J.~M.} \bibnamefont{Gilligan}},
  \bibinfo{author}{\bibfnamefont{W.~M.} \bibnamefont{Itano}},
  \bibinfo{author}{\bibfnamefont{D.~J.} \bibnamefont{Wineland}},
  \bibnamefont{and} \bibinfo{author}{\bibfnamefont{M.~G.}
  \bibnamefont{Raizen}}, \bibinfo{journal}{Phys. Rev. Lett.}
  \textbf{\bibinfo{volume}{70}}, \bibinfo{pages}{2359} (\bibinfo{year}{1993}).

\bibitem[{\citenamefont{Zhou et~al.}(2005)\citenamefont{Zhou, Zhang, and
  Guo}}]{ZhouXiang2005}
\bibinfo{author}{\bibfnamefont{X.-F.} \bibnamefont{Zhou}},
  \bibinfo{author}{\bibfnamefont{Y.-S.} \bibnamefont{Zhang}}, \bibnamefont{and}
  \bibinfo{author}{\bibfnamefont{G.-C.} \bibnamefont{Guo}},
  \bibinfo{journal}{Physical Review A} \textbf{\bibinfo{volume}{71}},
  \bibinfo{pages}{064302} (\bibinfo{year}{2005}), \bibinfo{note}{copyright (C)
  2009 The American Physical Society Please report any problems to
  prola@aps.org PRA}.

\bibitem[{\citenamefont{Hughes and Kamada}(2004)}]{HughesKamada2004}
    \bibinfo{author}{\bibfnamefont{S.}~\bibnamefont{Hughes}} \bibnamefont{and}
  \bibinfo{author}{\bibfnamefont{H.}~\bibnamefont{Kamada}},
  \bibinfo{journal}{Physical Review B} \textbf{\bibinfo{volume}{70}},
  \bibinfo{pages}{195313} (\bibinfo{year}{2004}), \bibinfo{note}{copyright (C)
  2009 The American Physical Society Please report any problems to
  prola@aps.org PRB}.

\bibitem[{\citenamefont{Englund et~al.}(2007)\citenamefont{Englund, Faraon,
  Fushman, Stoltz, Petroff, and Vuckovic}}]{EnglundFaraon07}
\bibinfo{author}{\bibfnamefont{D.}~\bibnamefont{Englund}},
  \bibinfo{author}{\bibfnamefont{A.}~\bibnamefont{Faraon}},
  \bibinfo{author}{\bibfnamefont{I.}~\bibnamefont{Fushman}},
  \bibinfo{author}{\bibfnamefont{N.}~\bibnamefont{Stoltz}},
  \bibinfo{author}{\bibfnamefont{P.}~\bibnamefont{Petroff}}, \bibnamefont{and}
  \bibinfo{author}{\bibfnamefont{J.}~\bibnamefont{Vuckovic}},
  \bibinfo{journal}{Nature} \textbf{\bibinfo{volume}{450}},
  \bibinfo{pages}{857} (\bibinfo{year}{2007}).

\bibitem[{\citenamefont{Srinivasan and Painter}(2007)}]{SrinivasanPainter2007}
    \bibinfo{author}{\bibfnamefont{K.}~\bibnamefont{Srinivasan}}
    \bibnamefont{and}
  \bibinfo{author}{\bibfnamefont{O.}~\bibnamefont{Painter}},
  \bibinfo{journal}{Nature} \textbf{\bibinfo{volume}{450}},
  \bibinfo{pages}{862} (\bibinfo{year}{2007}),
  \bibinfo{note}{10.1038/nature06274}.

\bibitem[{\citenamefont{Dutt et~al.}(2005)\citenamefont{Dutt, Cheng, Li, Xu,
  Li, Berman, Steel, Bracker, Gammon, Economou et~al.}}]{DuttCheng2005}
\bibinfo{author}{\bibfnamefont{M.~V.~G.} \bibnamefont{Dutt}},
  \bibinfo{author}{\bibfnamefont{J.}~\bibnamefont{Cheng}},
  \bibinfo{author}{\bibfnamefont{B.}~\bibnamefont{Li}},
  \bibinfo{author}{\bibfnamefont{X.}~\bibnamefont{Xu}},
  \bibinfo{author}{\bibfnamefont{X.}~\bibnamefont{Li}},
  \bibinfo{author}{\bibfnamefont{P.~R.} \bibnamefont{Berman}},
  \bibinfo{author}{\bibfnamefont{D.~G.} \bibnamefont{Steel}},
  \bibinfo{author}{\bibfnamefont{A.~S.} \bibnamefont{Bracker}},
  \bibinfo{author}{\bibfnamefont{D.}~\bibnamefont{Gammon}},
  \bibinfo{author}{\bibfnamefont{S.~E.} \bibnamefont{Economou}},
  \bibnamefont{et~al.}, \bibinfo{journal}{Physical Review Letters}
  \textbf{\bibinfo{volume}{94}}, \bibinfo{pages}{227403}
  (\bibinfo{year}{2005}).

\bibitem[{\citenamefont{Xu et~al.}(2009)\citenamefont{Xu, Yao, Sun, Steel,
  Bracker, Gammon, and Sham}}]{XuYao2009}
\bibinfo{author}{\bibfnamefont{X.}~\bibnamefont{Xu}},
  \bibinfo{author}{\bibfnamefont{W.}~\bibnamefont{Yao}},
  \bibinfo{author}{\bibfnamefont{B.}~\bibnamefont{Sun}},
  \bibinfo{author}{\bibfnamefont{D.~G.} \bibnamefont{Steel}},
  \bibinfo{author}{\bibfnamefont{A.~S.} \bibnamefont{Bracker}},
  \bibinfo{author}{\bibfnamefont{D.}~\bibnamefont{Gammon}}, \bibnamefont{and}
  \bibinfo{author}{\bibfnamefont{L.~J.} \bibnamefont{Sham}},
  \bibinfo{journal}{Nature} \textbf{\bibinfo{volume}{459}},
  \bibinfo{pages}{1105} (\bibinfo{year}{2009}),
  \bibinfo{note}{10.1038/nature08120}.

\bibitem[{\citenamefont{Bracker et~al.}(2005)\citenamefont{Bracker, Stinaff,
  Gammon, Ware, Tischler, Shabaev, Efros, Park, Gershoni, Korenev
  et~al.}}]{PhysRevLett.94.047402}
\bibinfo{author}{\bibfnamefont{A.~S.} \bibnamefont{Bracker}},
  \bibinfo{author}{\bibfnamefont{E.~A.} \bibnamefont{Stinaff}},
  \bibinfo{author}{\bibfnamefont{D.}~\bibnamefont{Gammon}},
  \bibinfo{author}{\bibfnamefont{M.~E.} \bibnamefont{Ware}},
  \bibinfo{author}{\bibfnamefont{J.~G.} \bibnamefont{Tischler}},
  \bibinfo{author}{\bibfnamefont{A.}~\bibnamefont{Shabaev}},
  \bibinfo{author}{\bibfnamefont{A.~L.} \bibnamefont{Efros}},
  \bibinfo{author}{\bibfnamefont{D.}~\bibnamefont{Park}},
  \bibinfo{author}{\bibfnamefont{D.}~\bibnamefont{Gershoni}},
  \bibinfo{author}{\bibfnamefont{V.~L.} \bibnamefont{Korenev}},
  \bibnamefont{et~al.}, \bibinfo{journal}{Phys. Rev. Lett.}
  \textbf{\bibinfo{volume}{94}}, \bibinfo{pages}{047402}
  (\bibinfo{year}{2005}).

\bibitem[{\citenamefont{Mermin}(1966)}]{Mermin1966}
    \bibinfo{author}{\bibfnamefont{N.~D.} \bibnamefont{Mermin}},
  \bibinfo{journal}{Journal of Mathematical Physics}
  \textbf{\bibinfo{volume}{7}}, \bibinfo{pages}{1038} (\bibinfo{year}{1966}).

\bibitem[{\citenamefont{Bateman et~al.}(1992)\citenamefont{Bateman, Bose,
  Dutta-Roy, and Bhattacharyya}}]{Bateman1992}
\bibinfo{author}{\bibfnamefont{D.~S.} \bibnamefont{Bateman}},
  \bibinfo{author}{\bibfnamefont{S.~K.} \bibnamefont{Bose}},
  \bibinfo{author}{\bibfnamefont{B.}~\bibnamefont{Dutta-Roy}},
  \bibnamefont{and}
  \bibinfo{author}{\bibfnamefont{M.}~\bibnamefont{Bhattacharyya}},
  \bibinfo{journal}{American Journal of Physics} \textbf{\bibinfo{volume}{60}},
  \bibinfo{pages}{829} (\bibinfo{year}{1992}).

\end{thebibliography}

\end{document}